\documentclass{article}
\usepackage{graphicx} 
\usepackage{fullpage}
\usepackage{amsthm}
\usepackage{amsmath}
\usepackage{float}
\usepackage{thmtools} 
\usepackage{subcaption}
\usepackage{multirow}
\usepackage{authblk}
\newtheorem{theorem}{Theorem}
\newtheorem{corollary}{Corollary}

\usepackage{xcolor}
\usepackage{colortbl}
\definecolor{jade}{rgb}{0.0, 0.66, 0.42}
\definecolor{cerise}{HTML}{CE4760}
\colorlet{fg}{jade!75!black}
\colorlet{bg}{cerise!75!black}
\colorlet{hl}{yellow!75!black}
\definecolor{mg}{RGB}{233,116,81}

\newcommand{\size}[1]{\lvert #1 \rvert}

\newcommand{\Domino}{Head-domino}
\newcommand{\domino}{head-domino}
\newcommand{\statecell}{\textbf{State-Cell}}
\newcommand{\SScycle}{280}
\newcommand{\SScycledir}{70}
\newcommand{\SSTC}{6}

\newcommand{\para}[1]{\paragraph{#1}} 

\usepackage{bbding}

\usepackage{rotating} 
\graphicspath{{./images/}}

\title{General Computation using Slidable Tiles with Deterministic Global Forces\thanks{This research was supported in part by National Science Foundation Grant CCF-2329918.}}

\author[1]{Alberto Avila-Jimenez}
\author[1]{David Barreda}
\author[1]{Sarah-Laurie Evans} 
\author[1]{Austin Luchsinger}
\author[1]{Aiden Massie} 
\author[1]{Robert Schweller} 
\author[1]{Evan Tomai}
\author[1]{Tim Wylie}
\affil[1]{University of Texas Rio Grande Valley}
\date{}

\begin{document}

\maketitle

\begin{abstract}
We study the computational power of the Full-Tilt model of motion planning, where slidable polyominos are moved maximally around a board by way of a sequence of directional ``tilts.'' We focus on the deterministic scenario in which the tilts constitute a repeated clockwise rotation.  We show that general-purpose computation is possible within this framework by providing a direct and efficient simulation of space-bounded Turing machines in which one computational step of the machine is simulated per $O(1)$ rotations.  We further show that the initial tape of the machine can be programmed by an initial tilt-sequence preceding the rotations.  This result immediately implies new PSPACE-completeness results for the well-studied problems of \emph{occupancy} (deciding if a given board location can be occupied by a tile), \emph{vacancy} (deciding if a location can be emptied), \emph{relocation} (deciding if a tile can be moved from one location to another), and \emph{reconfiguration} (can a given board configuration be reconfigured into a second given configuration) that hold even for deterministically repeating tilt cycles such as rotations. All of our PSPACE-completeness results hold even when there is only a single domino in the system beyond singleton tiles. Following, we show that these results work in the Single-Step tilt model for larger constant cycles. We then investigate computational efficiency by showing a modification to implement a two-tape Turing machine in the Full-Tilt model and Systolic Arrays in the Single-Step model. Finally, we show a cyclic implementation for tilt-efficient Threshold Circuits.

\end{abstract}

\section{Introduction}
The ``tilt'' motion-planning model~\cite{Becker2013MMCLPSR} is an elegant and simple model of multi-agent motion planning where polyominos are moved based on external global controls such as gravity, magnetic fields, or light.  The model consists of a 2D grid board with a mix of fixed immovable walls and a collection of slidable polyominos that move maximally based on directional \emph{tilt} commands.  By applying sequences of such directional tilts, polyominos can be rearranged into different permutations, possibly simulating computation.  If glues are added, the initial polyominos can combine into larger shapes, and the tilt framework serves as a model of self-assembly.  In this paper, we focus on one of the simplest variants of this model where the tilt sequence is a simple repeated clockwise rotation (a \emph{rotational} tilt sequence).

\para{Motivation and Previous Work.} The study of Tilt is motivated by the fact that this extremely simple and natural model of global control has extensive depth.  For example, in terms of computation, dual-rail logic devices and binary counters have been designed~\cite {Becker2019}.  In terms of complexity, PSPACE-completeness has been shown for various natural problems such as finding the minimum number of tilts to reconfigure a board~\cite{Becker2019}, or deciding if a configuration is reachable~\cite{Balanza:2020:SODA,FullTilt2019} at all.  In terms of self-assembly,
work has been done on designing boards to efficiently self-assemble target shapes in a pipelined manner~\cite{Becker2020}, in parallel~\cite{SML:2018:EPS}, in 3D~\cite{Keller2022}, and in a \emph{universal} manner where a single board can be programmed by a tilt sequence to construct a selected shape or pattern from a general class~\cite{Balanza:2020:SODA,FullTilt2019}.  Despite its power, the simplicity of global control permits applications at the macro, micro, and nano scale, including examples of large population robot swarms ranging from naturally occurring magnetotactic bacteria \cite{magnetic-bacteria2011,magnetic-bacteria2012,magnetic-bacteria2014} to manufactured light-driven nanocars \cite{Nanocars2012,Nanocars2005}.

\subsection{Our Focus and Results} 
Our focus is on a limited version of Tilt models in which the tilt sequences are a fixed, deterministic cycle of commands. Specifically, we utilize what may be the simplest such sequence:  a repeated clockwise rotation.  Previous complexity results no longer apply in this scenario, and the complexity and power of this limited variant of the model has remained open until now. We present three main results: the efficient simulation of space-bounded Turing machines, the PSPACE-completeness of classic problems in the model, and the efficient simulation of threshold circuits. An overview of these results is presented in Tables \ref{table:complexity} and \ref{table:computation}.

\para{Turing Machine Simulation.} 
Our first result is an efficient simulation of any given space-bounded Turing machine using a rotational tilt sequence (Section~\ref{sec:Turing}).  The size of the board is polynomial in the size of tape and the states of the Turing machine, and simulates one computational step every $O(1)$ rotations.  We further show that the tape of the Turing machine can be programmed by a preliminary tilt-sequence preceding the rotational sequence used to run the machine.  While previous work has implemented key components for computation within deterministic tilt-cycles (e.g., dual-rail logic and binary counters~\cite{Becker2019}), this is the first result to show that the fullest possible computational power of this model variant is realizable.  The direct and efficient simulation of Turing machines further provides a plausible framework for rotational Tilt systems to be implemented as general-purpose computational devices. 

\begin{table}[t]\small
    \centering \renewcommand{\arraystretch}{1.2}
    \begin{tabular}{|@{}c@{}|@{}c@{}|c|@{}c@{}|c|c|@{}c@{}|@{}c@{}||c|} \hline 
        Model&Tilt&Relocation&Occupancy&Vacancy&Reconfig.&Poly.&Poly.$>1$ & Ref.\\
        &Seq.&&&&&Size &Count &\\ \hline\hline
         \multirow{2}{*}{FT} & Det. & \textbf{PSPACE-C} & \textbf{PSPACE-C}& \textbf{PSPACE-C}& \textbf{PSPACE-C} &2&$1$ &Thm.~\ref{thm:RelRecOccVac} \\ \cline{2-9}
          & General & PSPACE-C & PSPACE-C & PSPACE-C$^*$ & PSPACE-C & 1 &0&\cite{Balanza:2020:SODA,Caballero:2020:CCCG1,FKR:2025:DFP}\\ \hline
        \multirow{2}{*}{SS} & Det. & \textbf{PSPACE-C} & \textbf{PSPACE-C}& \textbf{PSPACE-C}& \textbf{PSPACE-C} &2&$1$ &Cor.~\ref{thm:SSRelRecOccVac}\\ \cline{2-9}
          & General & PSPACE-C & PSPACE-C & PSPACE-C$^*$ & PSPACE-C & 1 &0&\cite{Balanza:2020:SODA,Caballero:2020:CCCG1,FKR:2025:DFP}\\ \hline
    \end{tabular}
    \caption{Summary of our results for the problems of relocation, occupancy, vacancy, and reconfiguration for deterministically repeating tilt cycles. The new results are in bold. The dash lines follow from the other results. ``Poly. Size'' is the maximum size of any polyomino, and ``Poly. $> 1$ Count'' is the number of polyominos in the system that are not singletons. $^*$This has not been formally proven, but was conjectured \cite{FKR:2025:DFP} and is likely true based off \cite{Caballero:2020:CCCG1}.}
    \label{table:complexity}
\end{table}

\begin{table}[t]\small
    \centering
    \renewcommand{\arraystretch}{1.2}
    \begin{tabular}{|c|c|c|c|c|c|c|c|c|} \cline{1-5}\cline{7-9}
    Model & Cycle & Single  & Multi- & Ref. & & Cycle & Threshold & Ref. \\ 
    & Length & Tape TM  & Tape TM &  & & Length & Circuits &  \\  \cline{1-5}\cline{7-9}\cline{1-5}\cline{7-9}
    
    Full-Tilt & 4 & $O(1)$  & $O(1)$ & Thms. \ref{thm:turingSim}, \ref{thm:2tape} & & 4 & $O(d)$ & Thm. \ref{thm:threshold}\\ \cline{1-5}\cline{7-9}
     
    Single-Step & \SScycle\ & $O(1)$  & $O(T)$ & Cors. \ref{thm:SSturingSim}, \ref{thm:SS2tape} & &\SSTC\ & $O(dw)$ & Cor.  \ref{thm:SSthreshold}\\ \cline{1-5}\cline{7-9}
    \end{tabular}
    \caption{Summary of deterministic computation results related to other models in terms of the cycles required for simulation. $T$ refers to the length of the tape of the Turing Machine. For circuits, $d$ and $w$ are the depth and width of the circuit, respectively. 
    }
    \label{table:computation}
\end{table}

\para{Complexity: Relocation, Occupancy, Vacancy, Reconfiguration.}
Our next set of results resolves the complexity of classic problems previously studied within the Tilt model (Section~\ref{sec:complexity}).  We focus on the problems of \emph{relocation} (deciding if a tile can be moved from one location to another), \emph{occupancy} (deciding if a given board location can be occupied by a tile), \emph{vacancy} (deciding if a location can be emptied), and \emph{reconfiguration} (can a given board configuration be reconfigured into a second given configuration).  In the case of general tilt sequences, these problems have been shown to be PSPACE-complete~\cite{Balanza:2020:SODA,FullTilt2019}, but their complexity for deterministic sequences was an open problem.  Our Turing machine simulation, along with special halting configurations, implies PSPACE-completeness for relocation, reconfiguration, occupancy, and vacancy even when there is only a single domino along with singletons in the system. 
A summary of our complexity results is given in Table~\ref{table:complexity}.

\para{Threshold Circuits.}
Along with Turing machines, circuits represent one of the most fundamental models of computation.  Previous work has shown how rotational Full-Tilt systems can simulate arbitrary circuits consisting of ANDs and ORs with a rotation count on the order of the circuit depth~\cite{Becker2019}.  We expand on this progress by adding the efficient simulation of the \emph{majority} gate that determines which bit value has the highest representation among a given set of input bits (Section~\ref{sec:tc}).  This extension yields the efficient simulation of \emph{Threshold} circuits, a provably faster set of circuits that is of particular interest to applications such as machine learning.


\subsection{Additional Related Work.}
Full-Tilt reconfiguration, even with almost no geometry, remains capable of non-trivial reconfiguration as studied in~\cite{ADK:2021:2048} with the game of 2048.  In terms of shape production, heuristic algorithms for planning the assembly of target shapes have been explored~\cite{BSB:2023:CMP}.  Closely related to the Full-Tilt model is the \emph{Single-Step} model of global control~\cite{Caballero:2020:CCCG1,Caballero2025} where each polyomino only moves a single step with each command, instead of maximally.  \cite{Caballero:2020:CCCG1} and \cite{FKR:2025:DFP} show that Single-Step can be the dual of Full-Tilt, implying a connection in the complexity between the two models for some classic problems.  But for some problems, such as shape production~\cite{akitaya2016trash,Akitaya:2022:FUN}, the problem is NP-hard even without any real geometry.  Other problems, such as relocation, are NP-hard even with only two possible directions for monotone~\cite{Caballero:2020:JIP}, or even no~\cite{Caballero2025}, geometry. The efficient assembly of shapes~\cite{Caballero:2021:UCNC_NC} and patterns~\cite{Caballero:2020:CCCG2} has also been explored.


\section{Preliminaries}\label{Prelims}

\textbf{Board.} 
A \emph{board} (or \emph{workspace}) is a rectangular region of the 2D square lattice in which specific locations are marked as \emph{blocked}.  Formally, an $m\times n$ board is a partition $B=(F,X)$ of $\{(x,y) | x\in \{1, 2, \dots, m\}, y\in \{1, 2, \dots, n\}\}$ where $F$ denotes a set of \emph{open} locations, and $X$ denotes a set of \emph{blocked} locations. We informally refer to blocked locations as ``concrete.'' 

\textbf{Tiles, Polyominos, and Configurations.}
A tile is a labeled unit square centered on a non-blocked point on a given board. Since we will not use any dynamic attachment between existing tiles, we simplify our definition. Formally, a tile is simply a coordinate $c=(x,y)$ on the board. A \emph{polyomino} is a finite set of non-overlapping tiles $p = \{t_1, \ldots, t_k\}$ that is connected with respect to the coordinates of the tiles.  A polyomino that consists of a single tile is informally referred to as a ``tile.''  A configuration $C=(B,P)$ is an arrangement of polyominos $P=\{p_1, \dots, p_k\}$ on a board $B$ such that there are no overlaps among polyominos, or with blocked board spaces.




\textbf{Step.} A \emph{step} is a way to turn one configuration into another by way of a global signal that moves all tiles/polyominos in a configuration one unit in a direction $\delta \in \{u,r,d,l\}$ (up, right, down, left) when possible without causing an overlap with a blocked position or another tile.  Formally, for a configuration $C=(B,P)$, let $P'$ be the maximal subset of $P$ such that translation of all polyominos in $P'$ by 1 unit in the direction $d$ induces no overlap with blocked squares or other polyominos. A step in direction $\delta$ is performed by executing the translation of all polyominos in $P'$ by 1 unit in that direction. If a configuration does not change under a step transition for direction $\delta$, we say the configuration is \emph{$\delta$-terminal}.  

\begin{figure}[t]
    \centering
	\begin{subfigure}[b]{0.12\textwidth}\centering
        \includegraphics[width=2.1cm]{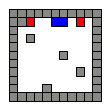}
        \caption{Initial}
    \end{subfigure}
	\begin{subfigure}[b]{0.12\textwidth}\centering
        \includegraphics[width=2.1cm]{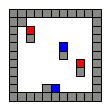}
        \caption{Down $\langle d\rangle$}
    \end{subfigure}
    \begin{subfigure}[b]{0.12\textwidth}\centering
        \includegraphics[width=2.1cm]{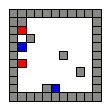}
        \caption{Left $\langle l\rangle$}
    \end{subfigure}
    \begin{subfigure}[b]{0.12\textwidth}\centering
        \includegraphics[width=2.1cm]{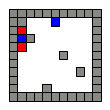}
        \caption{Up $\langle u\rangle$}
    \end{subfigure}
    \begin{subfigure}[b]{0.12\textwidth}\centering
        \includegraphics[width=2.1cm]{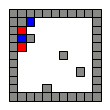}
        \caption{Left $\langle l\rangle$}
    \end{subfigure}
    \begin{subfigure}[b]{0.12\textwidth}\centering
        \includegraphics[width=2.1cm]{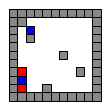}
        \caption{Down $\langle d\rangle$}
    \end{subfigure}
    \begin{subfigure}[b]{0.12\textwidth}\centering
        \includegraphics[width=2.1cm]{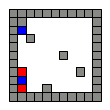}
        \caption{Left $\langle l\rangle$}
    \end{subfigure}
    \begin{subfigure}[b]{0.115\textwidth}\centering
        \includegraphics[width=2.1cm]{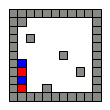}
        \caption{Down $\langle d\rangle$}
    \end{subfigure}
    \caption{Full-Tilt example of tiles moved through several tilts.}
    \label{fig:simple_example}
\end{figure}

\textbf{Tilt.} A \emph{tilt} in direction $\delta \in \{u,r,d,l\}$ for a configuration is executed by repeatedly applying a step in direction $\delta \in \{u,r,d,l\}$ until a $\delta$-terminal configuration is reached.  We say that a configuration $C$ can be \emph{reconfigured in one move} into configuration $C'$ (denoted $C \rightarrow_1 C'$) if applying one tilt in some direction $\delta$ to $C$ results in $C'$.  We define the relation $\rightarrow_*$ to be the transitive closure of $\rightarrow_1$. Therefore, $C \rightarrow_* C'$
means that $C$ can be reconfigured into $C'$ through a sequence of tilts.

\textbf{Tilt Sequence.} A \emph{tilt sequence} is a series of tilts inferred from a series of directions $D = \langle d_1, d_2,\dots, d_k \rangle$; each $d_i \in D$ implies a tilt in that direction. For simplicity, when discussing a tilt sequence, we just refer to the series of directions from which that sequence was derived. Given a starting configuration, a tilt sequence corresponds to a sequence of configurations based on the tilt transformation. An example tilt sequence $\langle d, l, u, l, d, l, d \rangle$ and the corresponding sequence of configurations can be seen in Figure~\ref{fig:simple_example}. 
A tilt sequence is termed \emph{deterministic} if it consists of repeated applications for a fixed constant-size cycle of tilt instructions. 
Our focus is a deterministic tilt sequence that is a \emph{rotational} sequence, which is the repeated application of the clockwise rotation $\langle u, r, d, l \rangle$.

\textbf{Turing Machine.} We use a Turing machine (TM) defined as a 7-tuple $(Q, \Sigma, \Gamma, \delta_U, q_0,$ $q_a,$ $q_r)$, where $Q$ is the set of states, and $\Sigma = \Gamma = \{0, 1\}$ for the tape and language alphabets, and $q_0, q_a, q_r$ are the start, accept, and reject states, respectively. For a universal machine, we use a 15-state, 2-symbol tape, as given in \cite{NW:2009:FI}, which defines the transition function $\delta_U$. We let $Q$ follow zero-based indexing. We assume a bounded tape of length $n$.



\section{Space-Bounded Turing Machine Simulation}\label{sec:Turing}

In this section, we detail how a rotational tilt system can efficiently simulate any space-bounded Turing machine.  The initial simulation is described in Section~\ref{sec:singletape}, and in Section~\ref{sec:turingProgram} we show how the Turing machine's initial tape can be programmed by a tilt-sequence that precedes the rotational sequence that runs the machine. Finally, we expand the construction to a two-tape Turing machine in Section~\ref{subsec:twotape}.
Our simulations are \emph{efficient} in that they simulate one step of the Turing machine per $O(1)$ rotations.  As an implication of our space-bounded Turing machine simulation, we show in Theorem~\ref{thm:RelRecOccVac} that the previously studied problems \cite{Balanza:2020:SODA,FullTilt2019,FKR:2025:DFP} of \emph{occupancy} (will a square on a board eventually get filled), \emph{vacancy} (will a square eventually be empty),  \emph{relocation} (can a given tile go from one location to another), and \emph{reconfiguration} (can all tiles be located to some set positions) are all PSPACE-complete even in the case of deterministic rotational sequences.  While each tilt application is defined as an application of single steps, for the purpose of these problems, we only care if the configuration is reached after a full application of the tilt, i.e., sliding through a target location during a tilt does not count as occupying that location. Here, we establish the following.

\begin{theorem}\label{thm:turingSim}
    For any Turing machine $\mathcal{M}$ with $s = |Q|$ states and a bounded tape of length $n$, there exists a non-bonding rotational Full-Tilt simulation of the machine with board size $O(ns^3)$ that simulates the machine at a rate of one step per $O(1)$ rotations.
\end{theorem}

\begin{figure}[t]
    \centering
    \includegraphics[width=.75\textwidth]{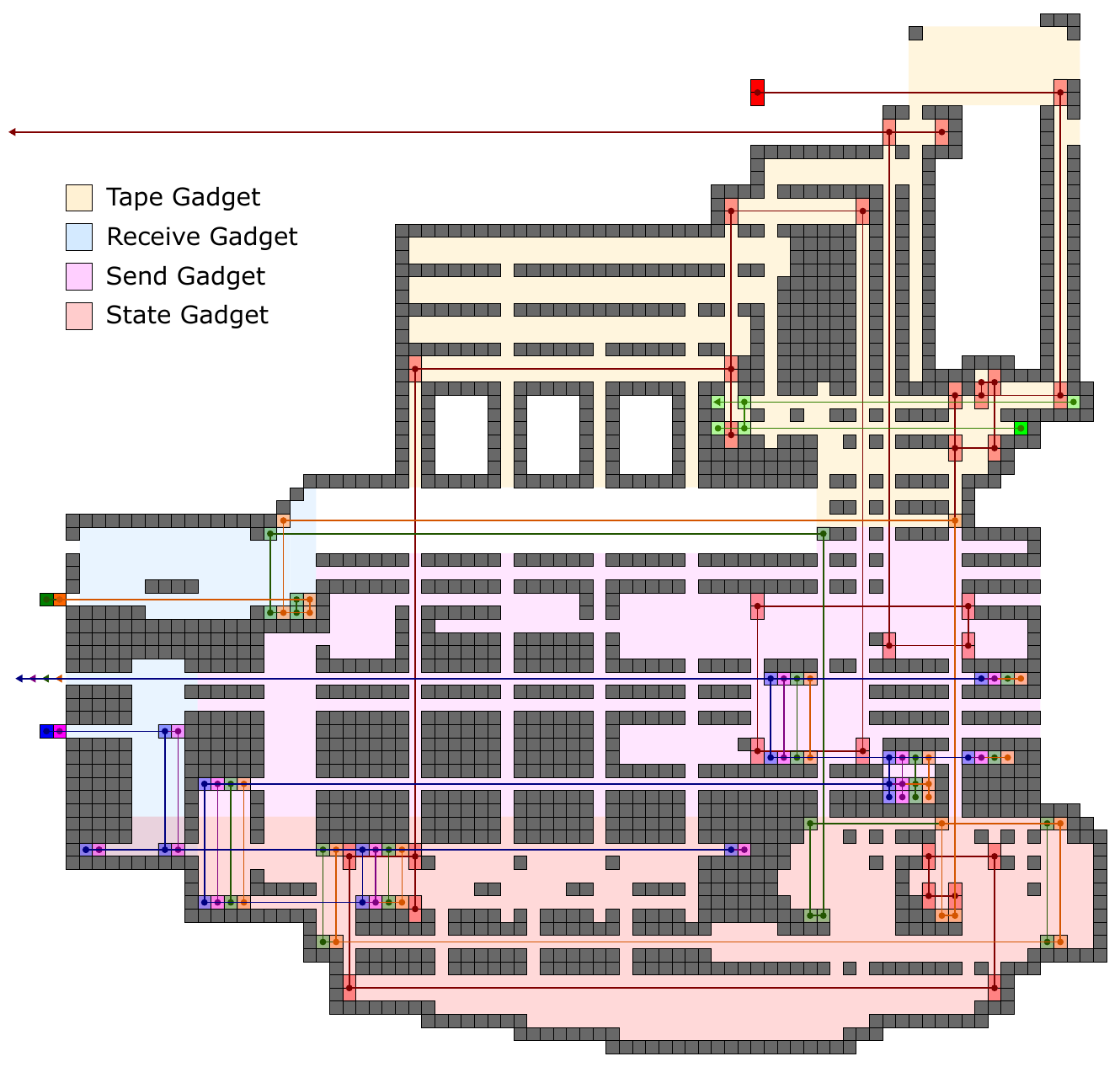}
    \caption{A \statecell\ construction (a single cell of the tape and the TM state control) with three states (including halt) performing an execution. The process starts with the \domino\ entering the \statecell\ from the left at the dark red domino at the top of the board. In the same cycle, at the leftmost blue and purple tiles, the {inert-singletons} enter. One cycle later, at the leftmost green and orange tiles, the {active-singletons} also enter. The \domino\ flows through the \emph{Tape gadget} reading $1$, \emph{State gadget} setting $q_0$, \emph{Tape gadget} writing $1$, then sends the $4$ {inert-singletons} left in the \emph{Send gadget} before finally being sent left itself. 
    }
    \label{fig:tm-cell-overview}
\end{figure}


By \emph{simulation}, we mean there exists a mapping from board configurations to bit-configurations of the Turing machine tape, with specific $O(1)$ size sub-regions of the board specifying these bit values.  A simulation then requires that the dynamics of the changing board configuration match the dynamics of the corresponding bit configuration for the Turing machine that is being simulated.  Our simulation board size is polynomial in the tape size and states of the Turing machine, uses $O(1)$ tilts per computation step, and is \emph{non-bonding}. 


\para{Design.} Typically, we envision a TM machine as a single control state with an infinite tape. In this model, our control structures are built from blocked locations (concrete) that are stationary, so the data must be sent from the current tape location to the control states. This is easily done in the Full-Tilt model, but we also need to keep track of the tape cell being used to send the data back. This is simple with two dominos (one to stay at the correct tape-cell location, and the other to handle state change), but is tricky with only a single domino. Further, no single-step implementation could be efficient since any given step might have an unbounded distance to traverse on the tape.

Here, we integrate the tape and control into a single structure where the control is repeated at every tape cell (overview in Figure \ref{fig:tm-cell-overview}). We can do this efficiently because a 2-symbol universal TM is possible with only 15 states \cite{NW:2009:FI}. Thus, the state control is constant size, and the controlling data is simply passed left or right with the head to process the data on the tape. This also allows for the single-step model to be universal for a constant cycle size since it now has a constant bounded distance to travel at any point.

\para{Programming.} From an implementation standpoint, programming a TM by placing particles on the tape could be difficult at a micro or nanoscale. We also discuss how the tape can easily be programmed by deviating from the deterministic cycle $O(n)$ or $O(\log n)$ times, where $n$ is the length of the tape. The linear programming is constant height, while the logarithmic method requires the tilt board to have $O(\log n)$ additional height.

\para{Extensions.} After presenting the single-tape machine, we show efficiency extensions. 
For Full-Tilt, we show how a 2-tape TM can be built with multiple dominos that are needed to communicate between the two tapes. For Single-Step, since the head and processing is local to the tape cell, we can extend the single-tape implementation to linear systolic arrays \cite{Brent:1984:ToC,Kung:1982:C,Mead:1980:BOOK} with a single domino for each tape head. 
The Full-Tilt model can also do systolic arrays on the single-tape TM, but the Single-Step model can not efficiently implement the 2-tape TM due to the distance required for the two tape heads 
to interact.

\begin{figure}[t]
    \centering
    \begin{subfigure}[b]{0.33\textwidth}\centering
        \includegraphics[width=1.\textwidth]{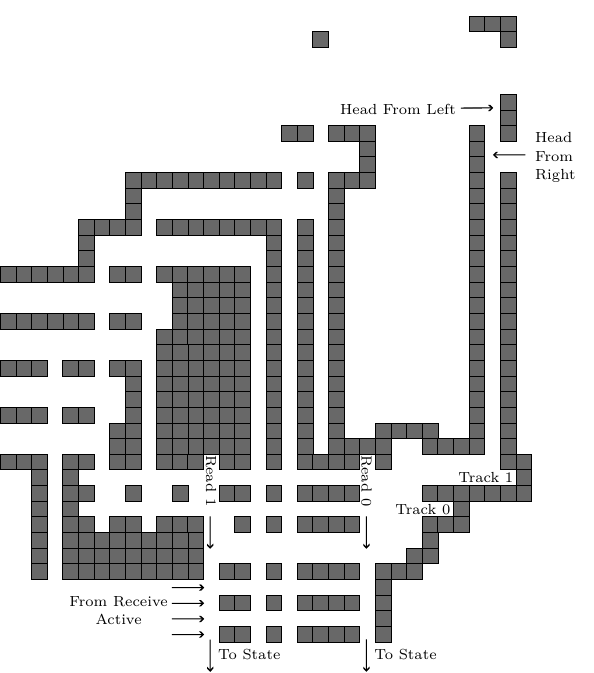}
        \caption{Tape Gadget with Labels}
        \label{fig:TReadingLabeled}
    \end{subfigure}
	\begin{subfigure}[b]{0.33\textwidth}\centering
        \includegraphics[width=.98\textwidth]{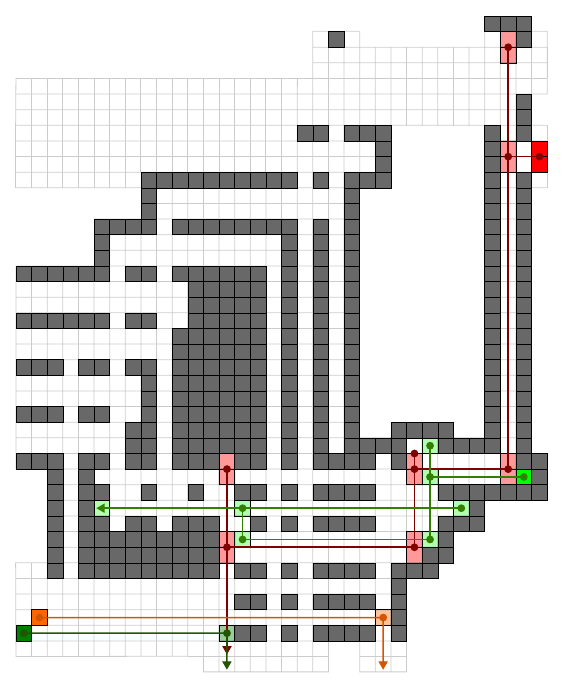}
        \caption{Read Operation (head on right)}\label{TapeRR1}
    \end{subfigure}
	\begin{subfigure}[b]{0.32\textwidth}\centering
        \includegraphics[width=1.\textwidth]{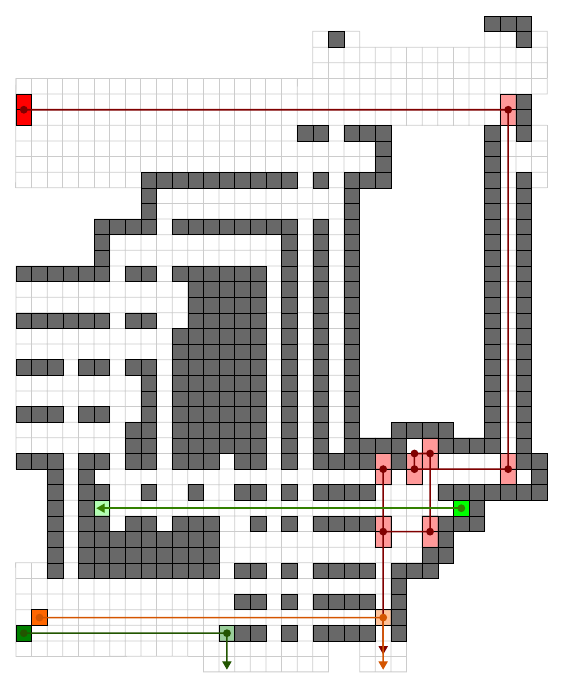}
        \caption{Read Operation. (head on left)}\label{TapeRL0}
    \end{subfigure}
    \caption{(a) The \emph{Tape gadget} construction, with labeled regions. (b-c) A \emph{read operation} performed in the \emph{Tape gadget}. Observe how the placement of the {data-singleton} affects where the \domino\ exits at the end of the operation.}
    \label{fig:TReading}
\end{figure}

\begin{figure}[t]
    \centering
    \begin{subfigure}[b]{0.32\textwidth}\centering
        \includegraphics[width=1.\textwidth]{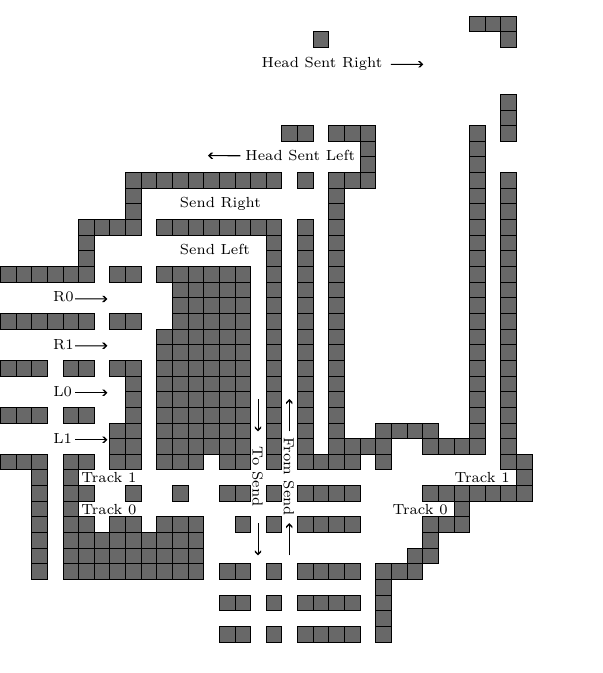}
        \caption{Labeled Tape}\label{fig:TapeWLabeled}
    \end{subfigure}
    \begin{subfigure}[b]{0.32\textwidth}\centering
        \includegraphics[width=1.\textwidth]{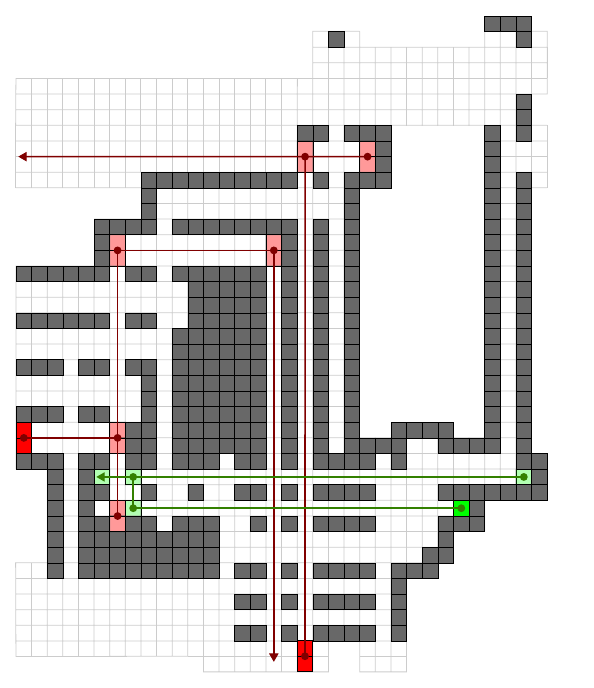}
        \caption{Write One, Move Left}\label{fig:TapeWEx1}
    \end{subfigure}
    \begin{subfigure}[b]{0.32\textwidth}\centering
        \includegraphics[width=1.\textwidth]{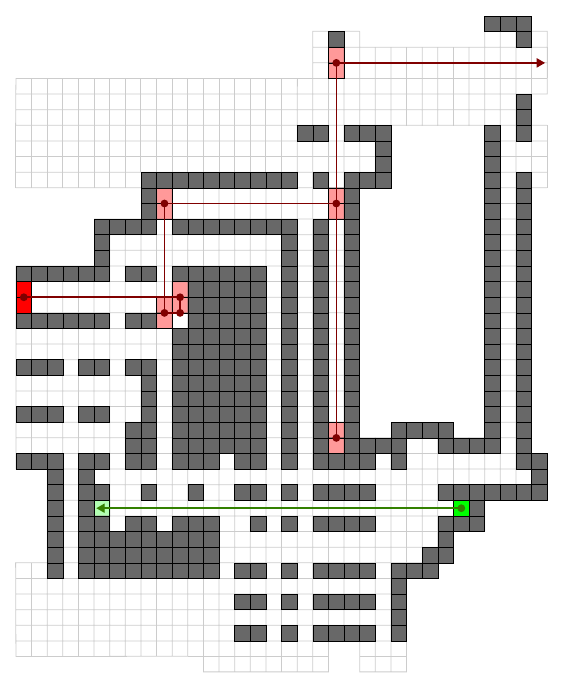}
        \caption{Write Zero, Move Right}\label{fig:TapeWEx2}
    \end{subfigure}
    \caption{(a) The \emph{Tape gadget} construction with the regions used for write labeled. (b) A write $L1$ operation performed in the \emph{Tape gadget}. The \domino\ moves the data-singleton from 0 to 1, then moves into the \emph{Send gadget}. It returns and is sent left. (c) A write $R0$ operation performed in the \emph{Tape gadget}. The \domino\ does not interfere with the data-singleton, then is sent right.}\label{fig:tapegadetall}
\end{figure}

\subsection{Simulation Construction}\label{sec:singletape}

Here, we provide the construction of our simulation of a single-tape space-bounded Turing machine.
Given a space-bounded Turing machine $\mathcal{M}=\{\mathcal{Q},\Sigma,\Gamma,\delta_U,q_0,q_a,q_r\}$ and an input tape $w$ of size $n$, we construct a board $B$ that simulates $\mathcal{M}$ over $w$. $B$ consists of $n$ copies of a \statecell\ that represents a specific tape cell on $w$ and a set of particles that model the behavior of $\mathcal{M}$.

\para{Particle Functions.} The set of particles consists of $2\size{\mathcal{Q}} + n$ singletons and one domino. They fall into a few categories based on function.

\begin{itemize}\setlength\itemsep{0em}
    \item \Domino. Implements the TM functionality and tracks the tape head location.
    \item {State-singletons}. 
    The singletons collectively represent $\mathcal{M}$'s current state based on the number of singletons used. At all points, the {state-singletons} are divided into \emph{active-singletons} and \emph{inert-singletons}. 
    
    \vspace{-.3cm}
        \begin{itemize}\setlength\itemsep{0em}
            \item {Active-singletons}. Tiles that collectively encode the current state of the Turing machine $\mathcal{M}$. Specifically, the number of {active-singletons} is equal to $2i$ where $q_i\in \mathcal{Q}$ is the current state of the Turing machine. 
            \item {Inert-singletons}. Any {state-singleton} that is not {active} is {inert}, and is stored in case the number of {active-singletons} needs to be increased. 
        \end{itemize}
    \vspace{-.2cm}
    \item {Data-singletons}. Each \emph{Tape gadget} has a singleton that is either in the `0' or `1' holding row for a dual rail implementation of the current symbol. 
\end{itemize}


\para{\statecell .} Each \statecell\ is composed of four gadgets: the \emph{Tape gadget}, \emph{State gadget}, \emph{Send gadget}, and \emph{Receive gadget}. See Figure \ref{fig:tm-cell-overview}.

\begin{itemize} \setlength\itemsep{0em}
    \item \emph{Tape Gadget} (Sec. \ref{ssec:tape_gadget}). The \emph{Tape gadget} holds the \statecell's {data-singleton}, whose location encodes the \statecell's symbol. The gadget's operations are Read and Write. Read sends the \domino\ to the same \statecell's \emph{State gadget} via one of two paths, which is determined by the location of the \emph{Tape gadget's} {data-singleton}. Write receives the \domino\ at one of four locations. The location determines where the {data-singleton} is placed, and which direction the {State-singletons} and \domino\ are sent. 
    
    
    \item \emph{State Gadget} (Sec. \ref{sssec:state-gadget}). The \emph{State gadget} serves the same purpose as the transition function $\delta_U$ of $\mathcal{M}$. It takes two pieces of information as its input: the holding row the \emph{Tape gadget's} {data-singleton} was in and the current count of the {active-singletons}. According to what $\delta_U$ maps the input to, the \emph{gadget} updates the {active-singleton} count and selects which path the \domino\ will return to the \emph{Tape gadget} on.
    
    \item \emph{Send Gadget} (Sec. \ref{sssec:send-gadget}). The \emph{Send gadget} moves the {state-singletons} out of the cell construct to the left or right, according to the position of the \domino.
    
    \item \emph{Receive Gadget} (Sec. \ref{sssec:receive-gadget}). The \emph{Receive gadget} catches the {state-singletons} from another cell construct and directs them to the \emph{Tape gadget} and \emph{State gadget}.

\end{itemize}

Together, these gadgets receive the \domino\ and {state-singletons} from another \statecell. They read the {data-singleton's} location, and based on what was read and the number of {active-singletons}, the gadgets update the count of the {active-singletons}, re-position the {data-singleton}, and move the \domino\ and {state-singletons} to an adjacent \statecell. 

\para{Geometric Alternating Cell Types.} 
\label{ssec:alternating}
The \statecell\ constructions are split into two types. Due to geometrical constraints, the location from which a \statecell\ sends the \domino\ to their right neighboring \statecell's \emph{Tape gadget} is to the left of the location where they will receive the \domino\ from their left neighboring \statecell's \emph{Tape gadget}. Because these locations serve different purposes, they cannot be at the same height. 
However, the location where a cell receives the \domino\ from the left must be vertically aligned with the location in the leftward adjacent cell from which the \domino\ is sent right. Therefore, \statecell\ constructions alternate between sending the \domino\ high and low. The send high constructions have the location from which they will send the \domino\ right placed above the location where they will receive \domino\ from the left. The send low constructions have the opposite design. Thus, by alternating the constructions, the location where the \domino\ is sent right in any given \statecell\ will be at the same height as the location its right neighbor receives the \domino\ from the left. This is also applied for the {active-singletons} and {inert-singletons} in the \emph{Send gadget}. All figures for this paper are of the \textsc{Send High}-type for simplicity.

\subsection{Gadgets}
We now discuss the gadgets in detail, and then walk through an execution cycle. 

\subsubsection{Tape Gadget}\label{ssec:tape_gadget}



\begin{figure}[t]
        \centering
    \includegraphics[width=.9\textwidth]{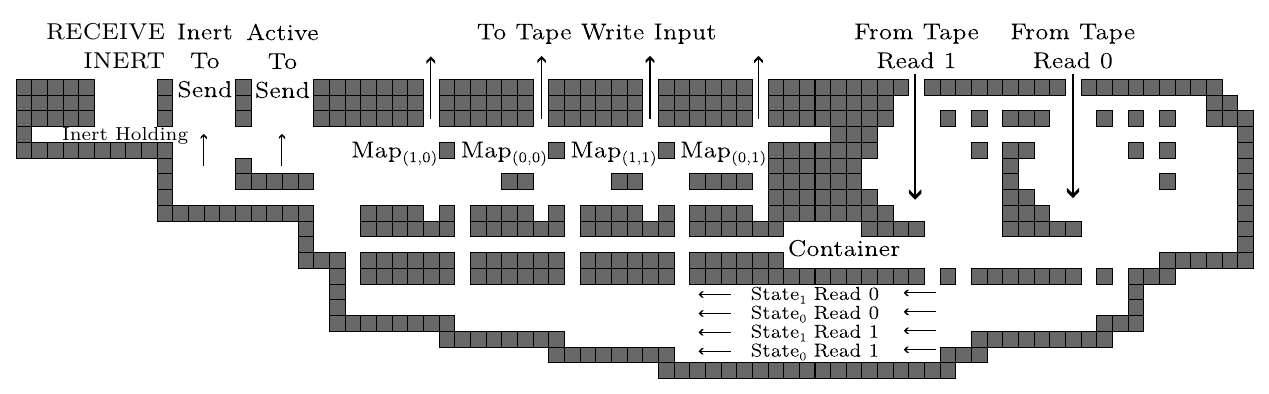}
    \caption{State Labeled. According to the State: $x$, and the Symbol read: $y$, the \domino\ is sent left at \textsc{State$_x$ Read $y$} (details in Figure \ref{fig:statecomp-interactions}). Or, if $q_x$ is a halt state, the \domino\ is sent into \textsc{container} (details in Figure \ref{fig:StateHalt}). Each \textsc{State$_x$ Read $y$} path leads to \textsc{Map$_{(x,y)}$} where the state is updated (details in Figure \ref{fig:stateup}). The new {active-singletons} and {inert-singletons} are sent to the \emph{Send gadget}, and the \domino\ is sent to one of the \emph{Tape gadget's} write inputs. }
\label{fig:StateLabeled}
\end{figure}


\para{Read Operation.}
The \emph{read operation} begins upon the \domino\ entering the \emph{Tape gadget}, shown in Figure \ref{fig:TReadingLabeled}, from another \statecell's \emph{Tape gadget}. Regardless of whether the \domino\ arrives at the \textsc{From Left} or \textsc{From Right} region, it traverses to \textsc{Track 1}. Then, according to the placement of the {data-singleton}, the \domino\ is placed at \textsc{Read 1} or \textsc{Read 0}. If the {data-singleton} is in \textsc{Track 1}, then it will interrupt the \domino's path and cause it to arrive at \textsc{Read 1}. In the process of interrupting, the {data-singleton} is moved to \textsc{Track 0}. Otherwise, the {singleton} is in \textsc{Track 0}, and the \domino\ will be placed at \textsc{Read 0} without affecting the {singleton}. In either case, after the \domino\ is at one of the \textsc{Read} locations, the {active-singletons} arrive from the \emph{Receive gadget} with half stopping under \textsc{Read 1}, and half under \textsc{Read 0}. Finally, the \domino\ will traverse down its chosen \textsc{Read} with half the {active-singletons} below it, while the rest of the {active-singletons} traverse down the other \textsc{Read}. Examples of the \emph{read operation} are shown in Figures \ref{fig:TReading}.

\para{Write Operation.}
The \emph{write operation} begins when the \domino\ is returned to the \emph{Tape gadget} from the same \statecell's \emph{State gadget}. The \domino\ returns at one of four locations: \textsc{$R0$}, \textsc{$R1$},  \textsc{$L0$}, or \textsc{$L1$}, shown in Figure \ref{fig:TapeWLabeled}. Each location maps to a direction-value combination, which dictates how the \domino\ moves and edits the {data-singleton}. For example, \textsc{$L1$} has the direction-value combination Left-1. If the write operation receives the \domino\ from that location, it writes 1 and moves the \domino\ left. 


Recall that after the read operation, the {data-singleton} is placed in \textsc{Track 0}. To write 1, the \domino\ blocks \textsc{Track 0}, sending the {data-singleton} to  \textsc{Track 1}. To write 0, because the read operation leaves 0 written, the \domino\ does not enter either of the \textsc{tracks} and instead waits one cycle. After writing, the \domino\ traverses to either \textsc{Move Left} or \textsc{Move Right}. \textsc{Move Left} sends the \domino\ into the \emph{Send gadget}, which will return it 2 cycles later. Upon returning, the \domino\ is sent to the left neighboring \statecell. \textsc{Move Right} delays the \domino\ once, then sends it to the right neighboring \statecell.

\begin{figure}[t]
        \centering
        \begin{subfigure}[b]{0.24\textwidth}\centering
            \includegraphics[width=1.\textwidth]{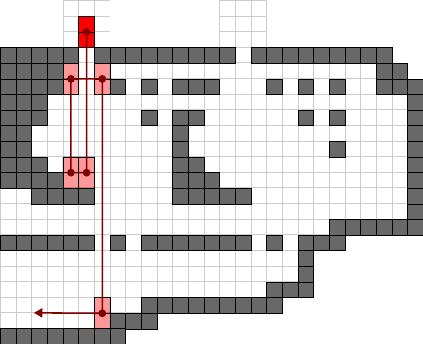}
            \caption{state 0, read 1}\label{fig:SC01}
        \end{subfigure}
        \begin{subfigure}[b]{0.24\textwidth}\centering
            \includegraphics[width=1.\textwidth]{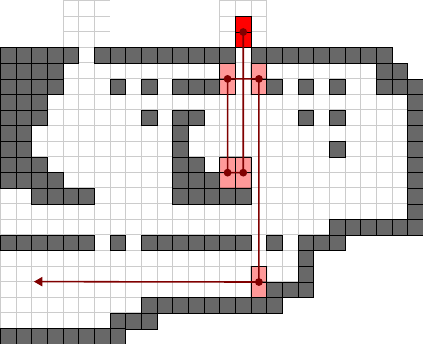}
            \caption{state 0, read 0}\label{fig:SC00}
        \end{subfigure}
        \begin{subfigure}[b]{0.24\textwidth}\centering
            \includegraphics[width=1.\textwidth]{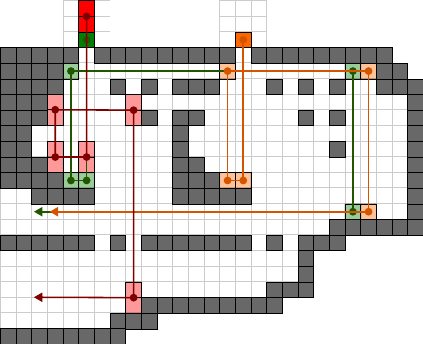}
            \caption{state 1, read 1}\label{fig:SC11}
        \end{subfigure}
        \begin{subfigure}[b]{0.24\textwidth}\centering
            \includegraphics[width=1.\textwidth]{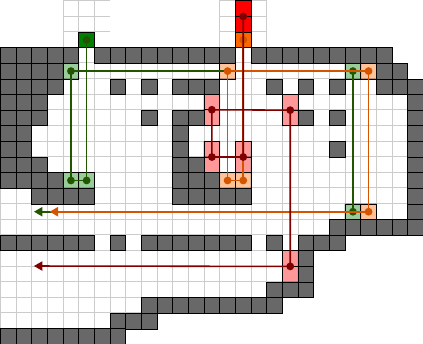}
            \caption{state 1, read 0}\label{fig:SC10}
        \end{subfigure}
        \caption{The four possible non-halt comparisons in the 3 state \emph{State gadget}. Depending on the initial location of the \domino\ and the count of {active-singletons}, different heights are traversed by the \domino. Recall that the number of {active-singletons} encodes the state: $q_x$ and the initial location of the \domino indicates the \statecell's symbol: $y$. Given that, the height traversed represents $(q_x,y)$, the input to $\mathcal{M}$'s transition function.}
        \label{fig:statecomp-interactions}
\end{figure}

\begin{figure}[t]
        \centering
        \begin{subfigure}[b]{0.45\textwidth}\centering
            \includegraphics[width=1.\textwidth]{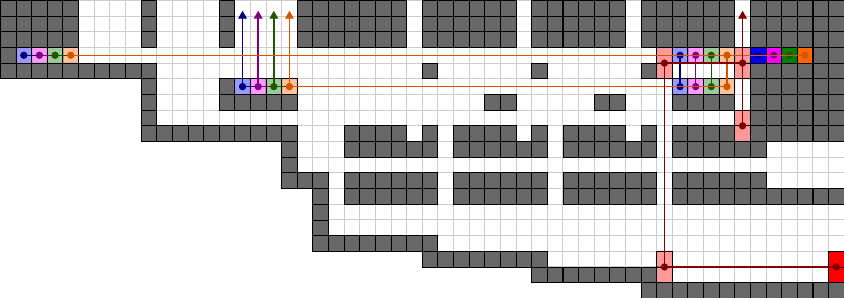}
            \caption{$(q_0,1)\rightarrow(q_2,1,R)$}\label{fig:SU01}
        \end{subfigure}
        \hspace*{.2cm}
        \begin{subfigure}[b]{0.45\textwidth}\centering
            \includegraphics[width=1.\textwidth]{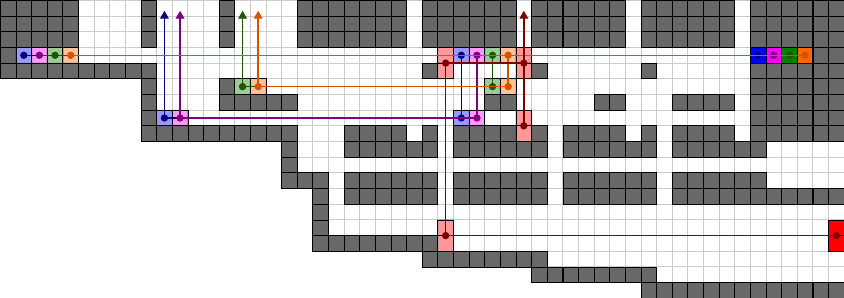}
            \caption{$(q_0,0)\rightarrow(q_1,0,R)$}\label{fig:SU00}
        \end{subfigure}
        \begin{subfigure}[b]{0.45\textwidth}\centering
            \includegraphics[width=1.\textwidth]{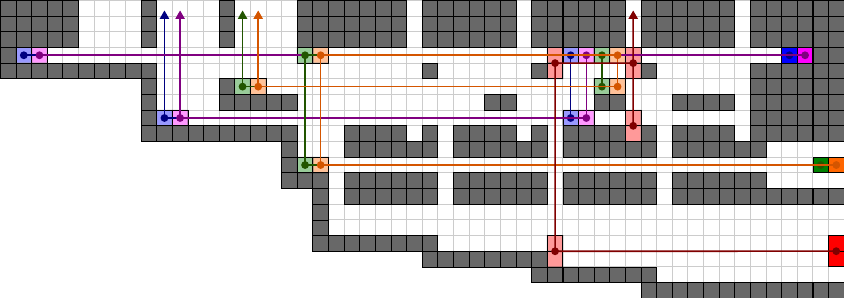}
            \caption{$(q_1,1)\rightarrow(q_1,0,L)$}\label{fig:SU11}
        \end{subfigure}
        \hspace*{.2cm}
        \begin{subfigure}[b]{0.45\textwidth}\centering
            \includegraphics[width=1.\textwidth]{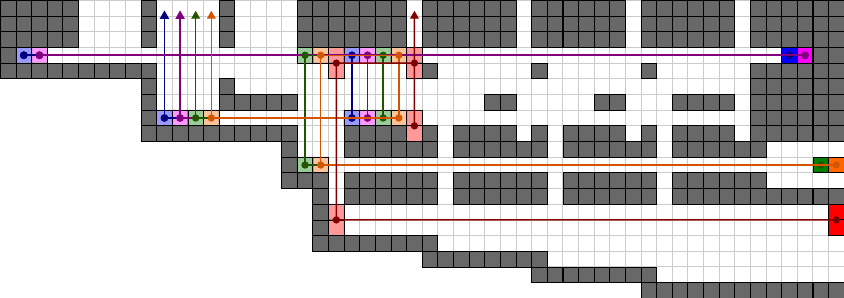}
            \caption{$(q_1,0)\rightarrow(q_0,1,L)$}\label{fig:SU10}
        \end{subfigure}
        \caption{The 4 possible non-halt \textsc{Map$_{(x,y)}$} options in a 3-state \emph{State gadget}. The \domino\ enters this section at a height indicating the input to $\mathcal{M}$'s transition function. Traversing from each height places the \domino\ above a unique \textsc{Map$_{(x,y)}$}. According to the transition function $(q_x,y)\rightarrow(q_a,b,c)$, the geometry under \textsc{Map$_{(x,y)}$} will set the number of {active-singletons} to $2 \times a$. (a) sets 4 {active-singletons} (b) sets 2 {active-singletons} (c) sets 2 {active-singletons} (d) sets 0 {active-singletons}. In addition, the tunnel the \domino\ exits \textsc{Map$_{(x,y)}$} from is connected to the \textsc{$cb$} input for the \emph{Tape gadget} write operation. If curious, in Figure \ref{fig:tm-cell-overview}, we can see the specific $cb$ tunnel connections. They are: (a) goes to \textsc{$R1$}, (b) goes to \textsc{$R0$}, (c) goes to \textsc{$L0$} and (d)  goes to \textsc{$L1$}.}
         \label{fig:stateup}
\end{figure}       
        

\begin{figure}[t]
        \centering
    \includegraphics[width=.9\textwidth]{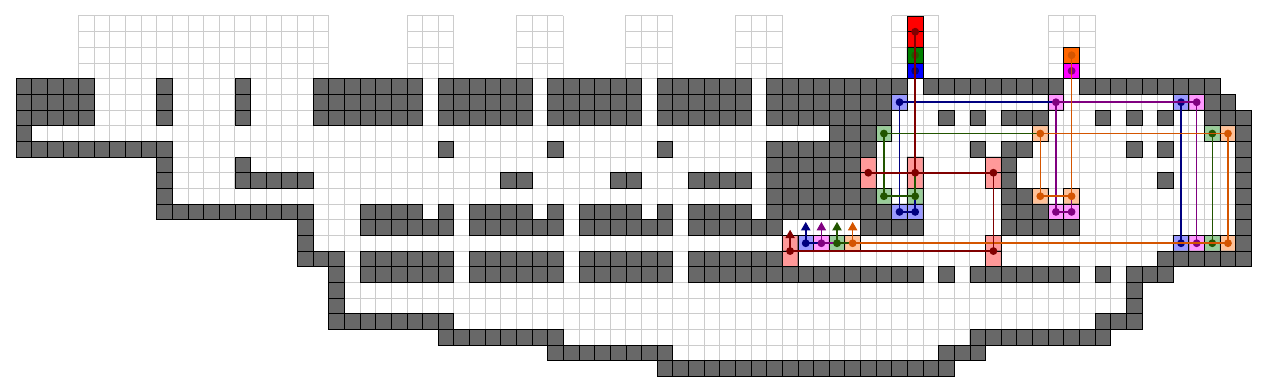}
    \caption{Simulating the Turing Machine Halting.
    The Halt States are encoded as all {state-singletons} being {active}. Instead of the \domino\ moving down to a \textsc{State$_x$ Read~$y$} combination, it stops one below the {active-singleton} line. Upon the following leftward tilt, the \domino\ catches the {active-singletons}, locking them, and itself, into the \textsc{container}.}
    \label{fig:StateHalt}
\end{figure}

\begin{figure}[t]
        \centering
    \includegraphics[width=.9\textwidth]{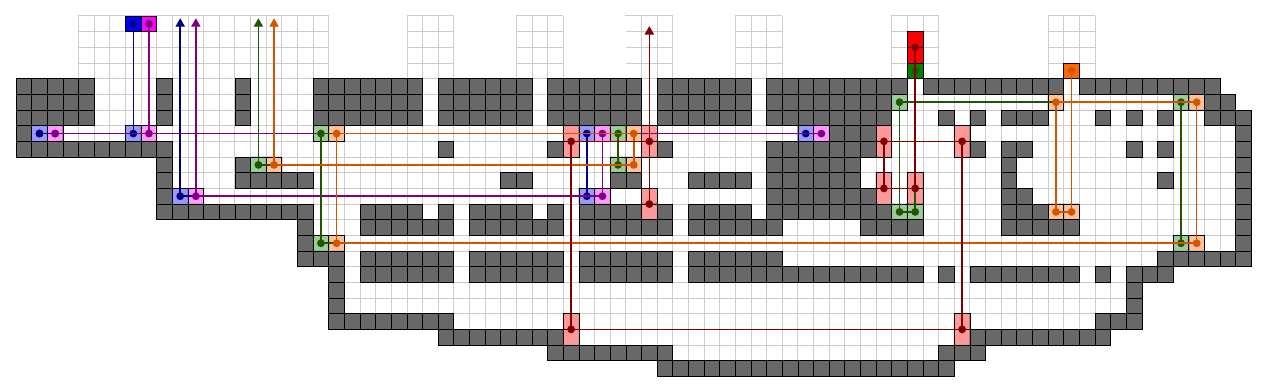}
    \caption{ Example of full \emph{State gadget} execution. The \domino\ comes \textsc{From Tape Read 1}. There is one {active-singleton} below it, indicating the simulation is in state $q_1$. The \domino\ goes to \textsc{State$_1$ Read 1}, and the two {active-singletons} join the two {inert-singletons}. The \domino\ moves up, blocking the {state-singletons} at \textsc{Map$_{(1,1)}$}. The Turing Machine's transition for $(1,1)$ is $(q_1,1)\rightarrow(q_1,0,L)$. In the \emph{State gadget}, we can see the {active-singleton} count being set to 2 according to that transition. For the same reason, the existing \domino\ traverses to \textsc{$L0$}.}
    \label{fig:StateChange}
\end{figure}

\subsubsection{State Gadget}\label{sssec:state-gadget}

  The \emph{State Gadget} receives half the {active-singletons} through \textsc{From Tape Read 1} and the other half through \textsc{From Tape Read 0} (Figure \ref{fig:StateLabeled}). In the same tilt, the \domino\ is placed atop one of the halves. The path the \domino\ takes forward is dependent on both the number of {active-singletons} offsetting it, which we define as $x$, and that \textsc{From Tape Read $y$} it entered through. If $q_x$ is one of the Turing machine's halt states, then the \domino\ stops at the height of the {active-singletons}. The next left tilt locks the {active-singletons} and the \domino\ in the container. An example is shown in Figure \ref{fig:StateHalt}. Otherwise, the \domino\ traverses left across the bottom section of the \emph{State gadget} at the height \textsc{State$_x$, Read $y$}. The height is set by the process in Figure \ref{fig:statecomp-interactions}. Meanwhile, the {active-singletons} are positioned to merge with the {inert singletons}, which have been cycling at \textsc{Wait}. Next, the geometry of \textsc{Map$_{x,y}$} sets a new count of {active-singletons} equal to $2\times a$ for $(q_x,y)\rightarrow(q_a,b,c)$ (see Figure \ref{fig:stateup}). The new {active} and {inert-singletons} then traverse separately to the \emph{Send gadget}, while the \domino\ is sent to \textsc{$R0$}, \textsc{$R1$}, \textsc{$L0$}, or \textsc{$L1$} (from Figure \ref{fig:TapeWLabeled}) according to which move-write combination $\delta_U$ maps $(q_x,y)$ to.



\begin{figure}[t]
        \centering
    \begin{subfigure}[b]{1.\textwidth}\centering
        \includegraphics[width=.8\textwidth]{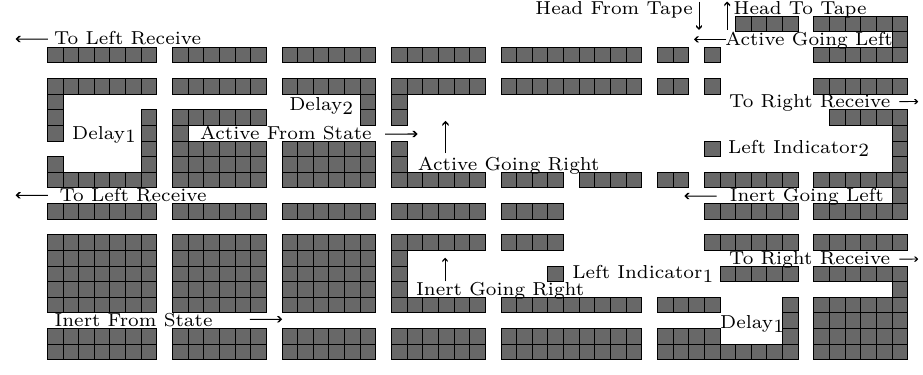}
        \caption{Send Labeled.} 
        \label{fig:SendLabeled}
    \end{subfigure}
    \begin{subfigure}[b]{0.45\textwidth}\centering
        \includegraphics[width=1.\textwidth]{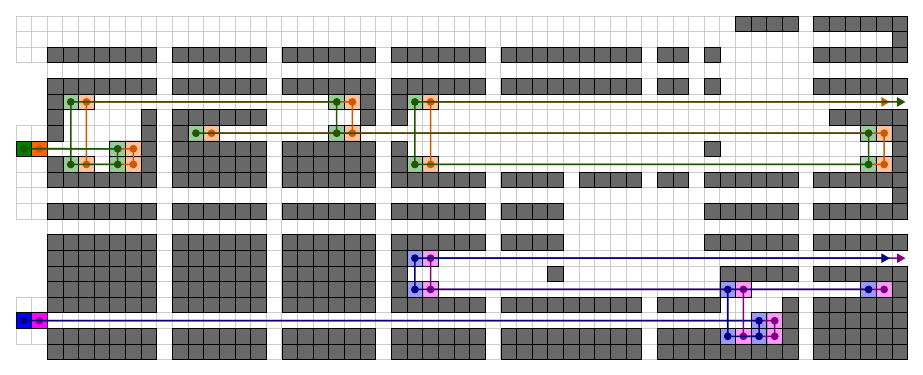}
        \caption{Send Right}
        \label{fig:SendR}
    \end{subfigure}
    \hspace{.3cm}
	\begin{subfigure}[b]{0.45\textwidth}\centering
        \includegraphics[width=1.\textwidth]{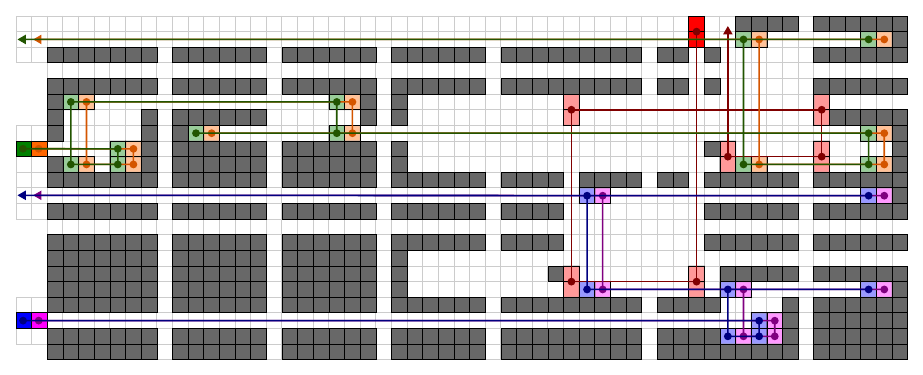}
        \caption{Send Left}
        \label{fig:SendL}
    \end{subfigure}
    \caption{(a) Labeled \emph{Send gadget}. The \emph{Send gadget} receives the state-singletons divided into Active and Inert. The inert-singletons are delayed once, then enter the lower chamber. At this point, there are two cases shown in (b) and (c). (b) shows the case where the \domino\ has not been sent by Tape write. This indicates \textsc{Move Right}. The inert-singletons pass \textsc{left indicator$_1$} and go to \textsc{Inert Going Right}. They exit \textsc{to Right Receive}. Active-singletons follow the same process in the upper chamber one cycle later. (c) shows the case where the \domino\ is sent down by the Tape write operation, triggering \textsc{Send Left}. Instead of moving past \textsc{Left Indicator$_1$} the \domino\ is caught on it. The inert-singletons are sent up to the \textsc{Inert Going Left} tunnel. While this happens, the \domino\ is sent up and cycles around to \textsc{Left Indicator$_2$}. The {active-singletons} are caught by it, behaving similarly to the {inert-singletons}. After Redirecting the {active-singletons}, the \domino\ is sent back into the \emph{Tape gadget} as in the bottom of Figure \ref{fig:TapeWEx1}.}\label{fig:sendgadgetall}
\end{figure}


\subsubsection{Send Gadget}\label{sssec:send-gadget}

Upon receiving the {active-singletons} and the {inert-singletons} from the \emph{State gadget}, the \emph{Send gadget} must delay them until the \emph{Tape gadget}'s write operation might send the \domino\ down. If the \emph{Tape gadget} received the \domino\ at \textsc{$L1$} or \textsc{$L0$}, then the \domino\ enters the \emph{Send gadget} to the right of \textsc{Left Indicator$_1$}. After the delay, if the \domino\ is present, the \emph{gadget} sends the {state-singletons} left via the process in Figure \ref{fig:SendL}. Otherwise, they are sent right, shown in \ref{fig:SendR}.



\begin{figure}[t]
    \centering
    \begin{subfigure}[b]{0.32\textwidth}\centering
        \includegraphics[width=1.\textwidth]{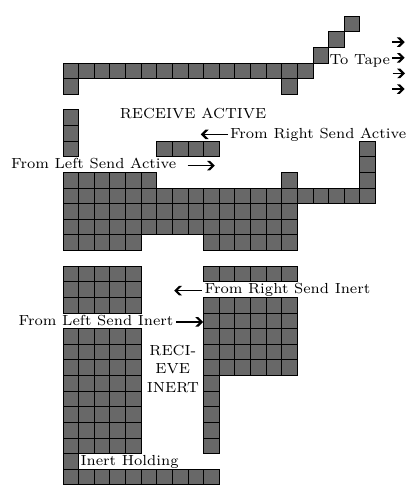}
        \caption{Receive Labeled}
        \label{fig:RecLabeled}
    \end{subfigure}
	\begin{subfigure}[b]{0.31\textwidth}\centering
        \includegraphics[width=1.\textwidth]{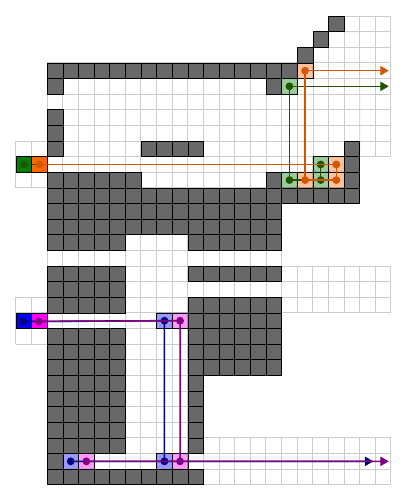}
        \caption{Left Receive}
        \label{fig:ReceiveL}
    \end{subfigure}
	\begin{subfigure}[b]{0.31\textwidth}\centering
        \includegraphics[width=1.\textwidth]{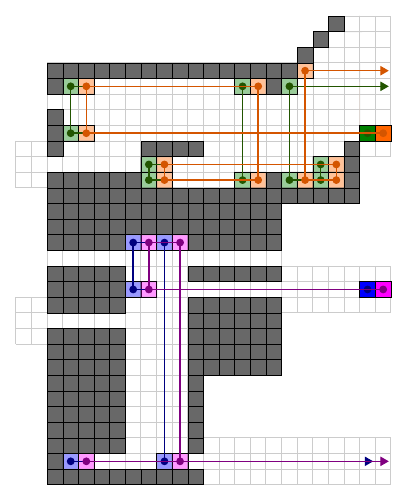}
        \caption{Right Receive}
        \label{fig:ReceiveR}
    \end{subfigure}
    \caption{(a) The \emph{Receive gadget} construction with labeled sections. This gadget catches the state-singletons (active and inert) from the adjacent cell to the right or left, then directs the active-singletons to the Tape Read input, and the inert-singletons to the \emph{State gadget} wait area. (b) Shows the gadget receiving two active and two inert-singletons from the Left. (c) Shows the gadget receiving two active-singletons and two inert-singletons from the Right.}
    \label{fig:Receive}
\end{figure}

\subsubsection{Receive Gadget}\label{sssec:receive-gadget}

The \emph{Receive gadget} directs the {active} and {inert-singletons} at the beginning of each execution. The {inert-singletons} are received from \textsc{Inert from Right} or \textsc{Inert from Left} as shown in Figure \ref{fig:RecLabeled}. They are moved to \textsc{Wait} in the \emph{State gadget} where they will cycle until accessed by the \domino. If the {active-singletons} are received from \textsc{Active from Left}, they are set to different heights, then sent to the \emph{Tape gadget}. If they are received from \textsc{Active from Right}, they are first delayed twice at \textsc{Delay} 1 and 2, then likewise have their heights changed and are sent to the \emph{Tape gadget}. The \textsc{from left} and \textsc{from right} cases are shown in Figures \ref{fig:ReceiveL} and \ref{fig:ReceiveR}, respectively.

\subsection{Execution Walkthrough}

Here, we walk through a single execution cycle shown in Figure \ref{fig:tm-cell-overview}, focusing on how the gadgets interact. Two {inert-singletons} enter the \emph{Receive gadget} at \textsc{From Left Send Inert}. The \domino\ enters the \emph{Tape gadget} at \textsc{Head From Left}, which begins the \emph{read operation}. One tilt cycle later, during the \emph{read operation}, the two {active-singletons} arrive at \textsc{From Left Send Active} in the \emph{Receive gadget}, and the {inert-singletons} are sent to \textsc{wait} in the \emph{State gadget}. The {active-singletons} are delayed one cycle to align with the \emph{read operation}, then are sent from the \emph{Receive gadget} to the \emph{Tape gadget}. At this point, the \emph{read operation} finishes, positioning the \domino\ in \textsc{Read 0} with one {active-singleton} below it, and the other below \textsc{Read 1}.

 The \domino\ and {active-singletons} traverse down to the \emph{State gadget}. The \emph{State gadget} receives as input: 1 {active-singletons} per \textsc{Tape Read}, and the \domino\ through \textsc{From Tape Read 0}. The Turing Machine's transition function $\delta_U$ dictates that $(q_1,0)\rightarrow(q_0,1,L)$. Thus, in the simulation, we see the \domino\ sent to \textsc{Map$_{(1,0)}$} where the {active-singleton} count is set to 0. The four {inert-singletons} are sent to the \emph{Send gadget}, and as $\delta_U$ orders, the \domino\ is sent to tunnel \textsc{$L1$} in the \emph{Tape gadget}.

In the \emph{Send gadget}, the {inert-singletons} are delayed. Because the \emph{Tape gadget} received the \domino\ from \textsc{$L1$}, it begins the \emph{write operation} by setting the {data-singleton} to 1. The \domino\ then traverses the \textsc{Move Left} tunnel to the \emph{Send gadget}. There, the {inert-singletons} have finished delaying, so they check for the \domino. Since it is present, they are redirected to \textsc{inert Going Left}, then are sent to the \statecell's left neighbor's \emph{Receive gadget}. Meanwhile, the \domino\ attempts to repeat the process of redirecting with the nonexistent \emph{active-singletons}, then traverses back up into the \emph{Tape gadget}, where it is sent to the \statecell's left neighbor's \emph{Tape gadget}.


\subsection{Complexity Results}\label{sec:complexity}

\begin{figure}[t]
    \centering
    \begin{subfigure}[b]{0.75\textwidth}\centering
        \includegraphics[width=1.\textwidth]{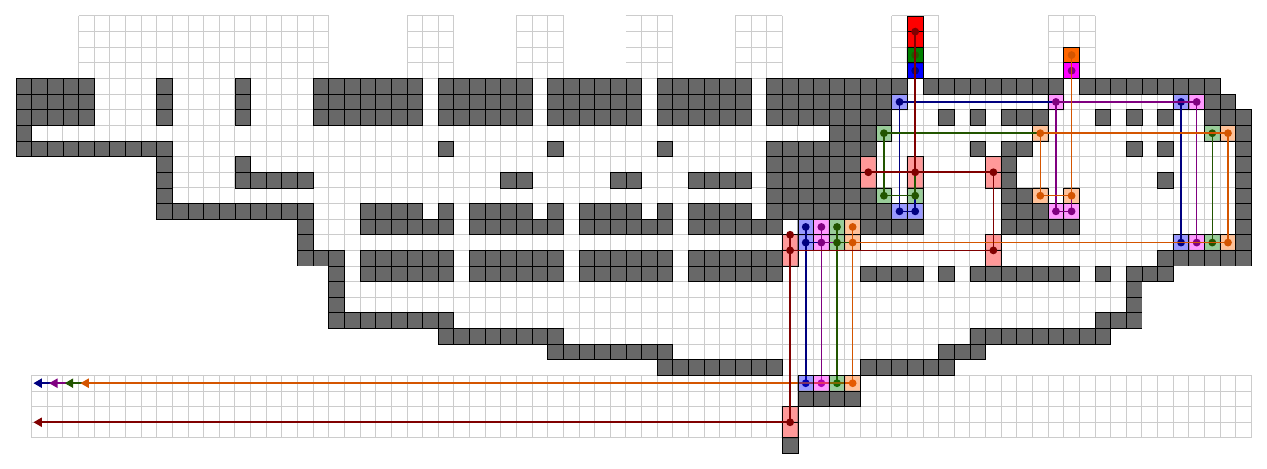}
        \caption{}\label{fig:ComplexityBase}
    \end{subfigure}
    \begin{subfigure}[b]{0.2\textwidth}\centering
        \includegraphics[width=.75\textwidth]{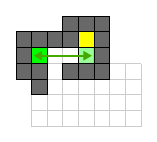}
        \caption{Vacancy Passive}\label{fig:VacancyP}
        \includegraphics[width=.75\textwidth]{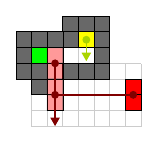}
        \caption{Vacancy Active} \label{fig:Vacancy}
    \end{subfigure}
    \caption{Updated State gadget for complexity results. (a) The modified \emph{State gadget} that abstracts out the cell at which a halt occurred. 
    (b) The behavior of the vacancy gadget when not receiving the \domino. The yellow tile can never move.
    (c) How vacancy is solved when the \domino\ is received. The \domino\ blocks the green tile, and the yellow tile can move down. 
    }\label{fig:updatedstateall}
\end{figure}




  

The simulation described above does not solve the occupancy, relocation, vacancy, or reconfiguration problems. This is because they deal with specific locations on the board. Our simulation has a clear indicator for the Turing machine's halt states, but the indicator could be at any of the $n$ \statecell\ constructions. We make a small modification to all \statecell s, such that they eject the state-singletons when the halt state is reached (Figure \ref{fig:ComplexityBase}).
From this and a few other small changes, we achieve the following results.


\begin{theorem}\label{thm:RelRecOccVac}
    Relocation, reconfiguration, occupancy, and vacancy are PSPACE-complete in the deterministic Full-Tilt model with rotational sequences (cycle-length of 4, using 4 directions), even for non-binding systems with singletons and a single polyomino of size 2.
\end{theorem}
\begin{proof}
Given Theorem \ref{thm:turingSim}, it is PSPACE-complete to know whether the halting state of any Turing machine encoded as a Full-Tilt system can be reached.  In this construction, the halting state causes the state-singletons to be ejected (Figure \ref{fig:ComplexityBase}), which we connect from all the \textsc{containers} from every \statecell\ to one location. This immediately proves relocation and occupancy can simulate a bounded Turing Machine, as after falling out of any container, the \domino\ is moved left to a predetermined location. Vacancy can be solved by attaching a \emph{Vacancy gadget} (Figure \ref{fig:VacancyP}) to that location such that the \domino\ follows the path in Figure \ref{fig:Vacancy}, allowing the {yellow-singleton} to move for the first time. 


Reconfiguration requires assignment of a target location for every particle. This includes the {data-singletons}. We solve this problem after reaching the halt state by setting every data-singleton to `1'. This is done by sending the \domino\ through mildly modified \emph{Tape gadgets}, as shown in Figure \ref{fig:ModdedTape}. It is passed to the leftmost cell's modified \emph{Tape gadget}  after exiting the containment area. The final configuration is all data-singletons set to `1', the state-singletons trapped in a cycle from the move after halt, and the \domino\ located above and to the right of the rightmost cell.
\end{proof}



\subsection{The Single-Step Model} \label{ssec:sstm}

Here, we note that the main complexity problems are still hard in the Single-Step model for a constant cycle length. This is achieved by adapting the Full-Tilt construction.
Then with the same modifications to the Full-Tilt TM construction for each specific problem, we can achieve the corresponding results for the complexity questions. 

\begin{corollary}\label{thm:SSturingSim}
    For any Turing machine $\mathcal{M}$ with $s = |Q|$ states and a bounded tape of length $n$, there exists a non-bonding rotational Single-Step simulation of the machine with board size $O(ns^3)$ that simulates the machine at a rate of one step per $O(1)$ rotations.
\end{corollary}


\begin{proof}
Given a TM $M$, we encode it to be used in a universal 15-state 2-symbol TM, and we create that universal machine using the method outlined in Theorem \ref{thm:turingSim} for Full-Tilt.
Since the size of the universal TM is constant, the maximum distance any tilt causes a tile to move within a cell is \SScycledir, and thus the same construction works with a cycle of $\langle u^{\SScycledir}, r^{\SScycledir}, d^{\SScycledir}, l^{\SScycledir} \rangle$. Only a slight modification is needed to move pieces between \statecell s. Every \SScycledir\ steps, a small holding area of simple geometry (see Figure \ref{fig:SSpath}) is needed. 
\end{proof}


\begin{corollary}\label{thm:SSRelRecOccVac}
    Relocation, reconfiguration, occupancy, and vacancy are PSPACE-complete in the deterministic Single-Step model with rotational sequences (cycle-length of \SScycle, using 4 directions), even for non-binding systems with singletons and a single polyomino of size-2. 
\end{corollary}


\begin{proof}
    Using the \statecell\ constructions and gadgets from Theorem \ref{thm:RelRecOccVac}. Since the maximum distance within any gadget is less than \SScycledir, they work the same as in Full-Tilt. We connect the \statecell\ constructions with Single-Step paths with \SScycledir-spaced holds (e.g., Fig. \ref{fig:SSpath}).
\end{proof}

We note that we can actually improve Corollary \ref{thm:SSRelRecOccVac} to show that the problems are still PSPACE-complete for a cycle length of 40 as $\langle u^{10}, r^{10}, d^{10}, l^{10} \rangle$, but the proof uses a different technique. It is proven by implementing gadgets in the motion planning framework \cite{DCL_MCU2022}. Due to the additional preliminaries and explanations that would be required, and for consistency and elegance of presentation, we have omitted this result in favor of the less optimal one that follows as a corollary.

\begin{figure}[t]
    \centering
    \begin{subfigure}[b]{0.6\textwidth}\centering
        \includegraphics[width=.97\linewidth]{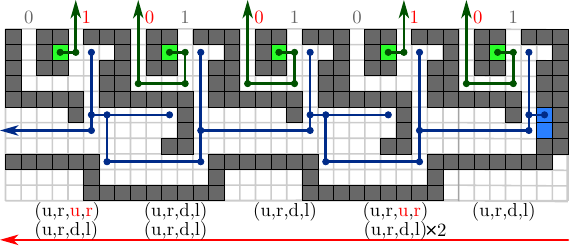}
        \caption{}\label{fig:TM-Domino-Prog}
    \end{subfigure}   
    \begin{subfigure}[b]{0.35\textwidth}\centering
        \includegraphics[width=1.\linewidth]{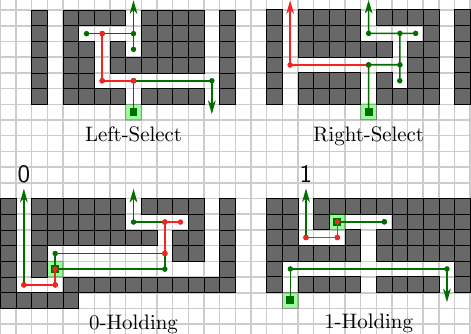}
        \caption{}\label{fig:TM-bit-selector}
    \end{subfigure}   
    \caption{(a) In Full-Tilt, selecting $k$ 1's in the tape in $O(n)$ time where $n$ is the length of the tape and with $O(k)$ edits to the cycle. The cycle ejects a tile in the `0' location, unless an additional $\langle u, r, u, r \rangle$ is added, which causes the tile to exit in the `1' location. After the tilts, the domino exits to the starting cell of the tape. (b)  Components of a bit-selection gadget.  The left and right-select gadgets dictate the selection sequences for the bit selectors.  The 0-holding gadget holds tiles that were not selected as 1's.  The 1-holding gadget keeps selected tiles in place until all tiles are ready to be extracted. The head domino is locked in place until a $\langle d,l,d,l \rangle$ sequence, which is the sequence to extract all selected tiles.}
    \label{fig:TM-Domino-Prog-bit-selector}
\end{figure}

\subsection{Programming the Tape}\label{sec:turingProgram}

Given that a 2-symbol universal Turing Machine can be made with as few as 15 states \cite{NW:2007:FSU}, and the applications at the nanoscale may make placing singletons tedious, there is good motivation to create a tape that is programmable at run time through the global external forces rather than by placing the tiles in the tape directly. However, for a deterministic system, there is no way to achieve this.  We provide two methods for initializing the tape:  the first being a simple linear encoding that requires $O(n)$ tilts for a size $n$ tape, and a second that uses specialized bit-selection gadgets that can place $k$ 1's on the tape (at any desired positions) with a $O(k \log n)$-length sequence. 

We make a slight modification to our deterministic system to allow a set number of modifications to the cycle at the beginning of the simulation. Once the ``programming'' is executed, the system goes into the standard $\langle u,r,d,l \rangle$ cycle and may not be modified again.  To ensure the tilt sequence modifications of the programming sequence do not interfere with the rest of the construction, we can 
assume 1) the \domino\ is stored in a holding cell and can only be extracted with a $\langle d,l,d,l \rangle$ sequence and 2) that the Turing machine starts in state zero with all state-singletons in the area labeled \textsc{Wait} of the \emph{State gadget} (Fig.~\ref{fig:StateLabeled}).

\para{Linear Edits.} In our Full-Tilt TM, a 0 (1) is represented by a single tile in \textsc{Track 0} (\textsc{Track 1}) of the \emph{Tape gadget} 
(Figure~\ref{fig:TapeWLabeled}).  Before  programming, we assume neither track contains a tile. 
An example of a linear programming process is shown in Figure \ref{fig:TM-Domino-Prog} where the $\langle u,r,d,l \rangle$ cycle walks a domino to each tape cell and assigns a tile to either Track 0 or Track 1.  To program a `1' in a tape cell, the $\langle u,r,d,l \rangle$ sequence only needs a slight edit by replacing the down and left commands 
to give $\langle u,r,u,r \rangle$.  Thus, programming the tape requires $O(k)$ edits where $k$ is the number of 1's on the tape and a total of $O(n)$ cycles where $n$ is the length of the tape.
The tiles being sent can be spaced correctly to be intercepted by extending the \textsc{Track 1} and \textsc{Track 0} tunnels to the right of the \emph{Tape gadget}.



\begin{figure}[t]
    \centering
    \includegraphics[width=1.\linewidth]{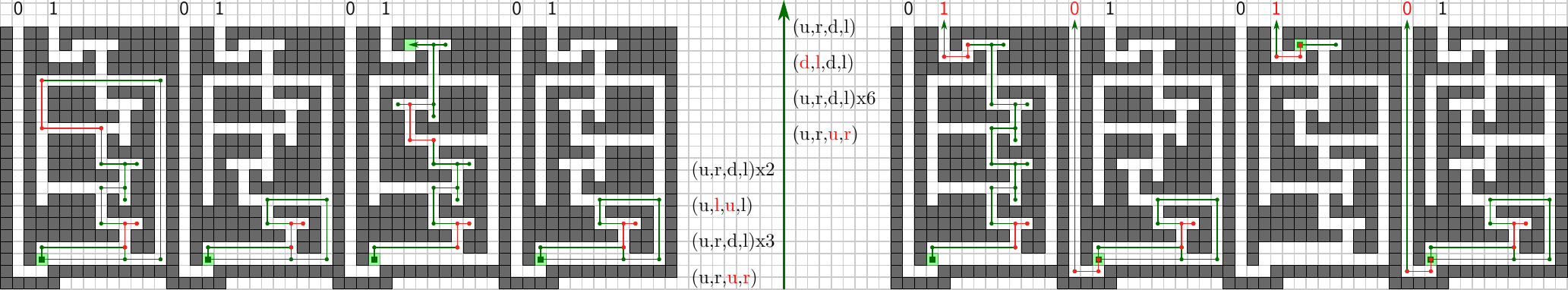}
    \caption{Full \emph{Bit-selector gadgets} for a 4-bit tape. (Left) A sequence of tilt rotations to flip the second bit to a 1. (Right) A subsequent sequence of tilt rotations to flip the fourth bit to a 1 and extract all four bits from the gadgets.  In general, $k$ many 1's can be selected for a length-$n$ tape in $O(k\log n$) tilts and with $O(k\log n$) edits to the cycle.}
    \label{fig:TM-bit-selector-full}
\end{figure}

\para{Logarithmic Tilts.}
We exploit the global control available and use \emph{Bit-selector gadgets} that allow us to select precisely which bits should become 1's without traversing the length of the tape, nor requiring the use of a domino. Figure~\ref{fig:TM-bit-selector} shows the components of these gadgets, and Figure~\ref{fig:TM-bit-selector-full} presents a full example for programming a 4-bit sequence.  The \emph{Bit-selector gadgets} are comprised of the left-select, right-select, 0-holding, and 1-holding gadgets.  Each gadget has a 0-holding component at its base (initialized with a tile in it), followed by a sequence of left- and right-select components, and is capped with a 1-holding component.
By using a unique sequence of left- and right-select components per bit, the gadget repeatedly eliminates half the bits per tilt cycle, until one bit is chosen to become a 1. 

To begin this process, we modify the first $\langle u,r,d,l \rangle$ rotation to become $\langle u,r,u,r\rangle$ in order to move all tiles through the top of the 0-holding component.  To advance through a left-select gadget, we use another modified rotation step of $\langle u,l,u,l \rangle$.  To advance through a right-select component, the standard $\langle u,r,d,l \rangle$ rotation is used.  After $O(\log n)$ of these choices, exactly one tile is in the 1-holding component.  All others have been redirected back to the 0-holding component through the side tunnels.  This process is repeated until all desired 1's have been selected. After this, a final modified tilt rotation of $\langle d,l,d,l \rangle$ prepares all bit tiles (0's and 1's) to be sent to the \emph{Tape gadget} at the same time.  Thus, we can program any $n$-bit tape with $k$ many 1's using $O(k \log n)$ tilt rotations with $O(k \log n)$ edits to the tilt cycle.  As before, the tiles exiting the selection gadgets can be positioned to intersect with the correct track of the \emph{Tape gadget}.
\section{Space-bounded Two-tape Turing Machine}\label{subsec:twotape}
We now show how the same \statecell\ construction from Section \ref{sec:singletape} can be extended to simulate a Turing machine with two tapes: $\alpha$ and $\beta$. By modifying the \emph{State gadget}, we are able to make new gadgets that have the head-dominos of 2 tapes interact and share information. This information is then sent back into each tape, which will accordingly execute a transition on one of their \statecell\ constructs, following the process described throughout Section \ref{sec:singletape}.

\begin{figure}[t]
    \centering
    \includegraphics[width=.8\linewidth]{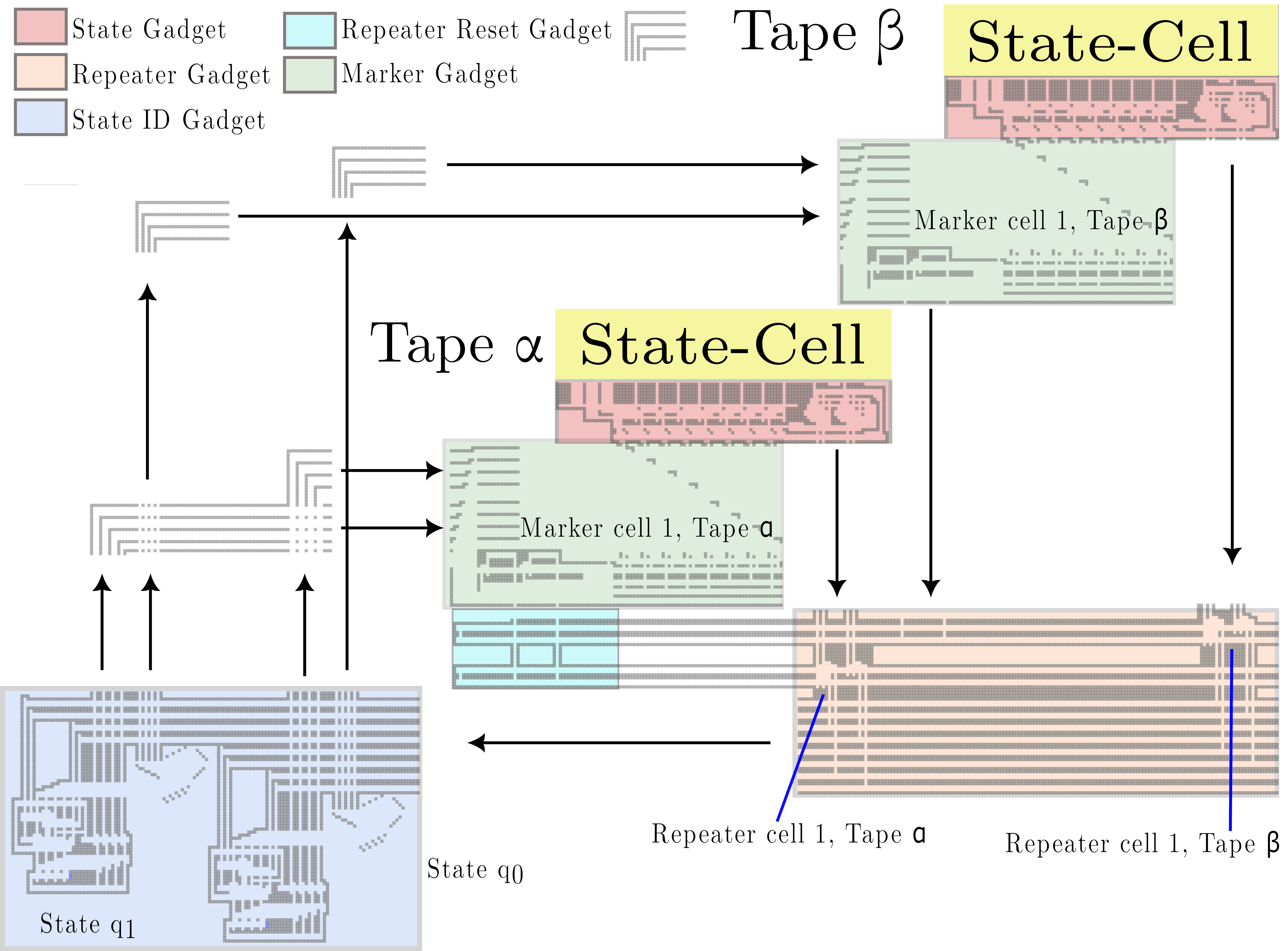}
    \caption{A full diagram of the first cells in a two-tape Turing machine. The yellow boxes indicate the upper half of the \statecell, which works the same as the one in Figure \ref{fig:tm-cell-overview}. The arrows show how the Turing machine processes information. The unshaded tunnels redirect any helper-dominos moving into the different gadgets.} 
    \label{fig:multitape-diagram}
\end{figure}

\para{Two-tape Helper Dominos.} In addition to the particles already defined in Section~\ref{sec:singletape}, we introduce helper-dominos used in the two-tape gadgets.

\vspace{-.4cm}

\begin{figure}[t]
        \centering
        \begin{subfigure}[b]{0.34\textwidth}\centering
            \includegraphics[height=3.6cm]{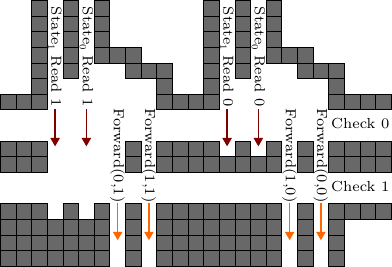}
            \caption{Repeater Gadget}\label{fig:repeater-labeled}
        \end{subfigure}
        \begin{subfigure}[b]{0.35\textwidth}\centering
            \includegraphics[height=5.9cm]{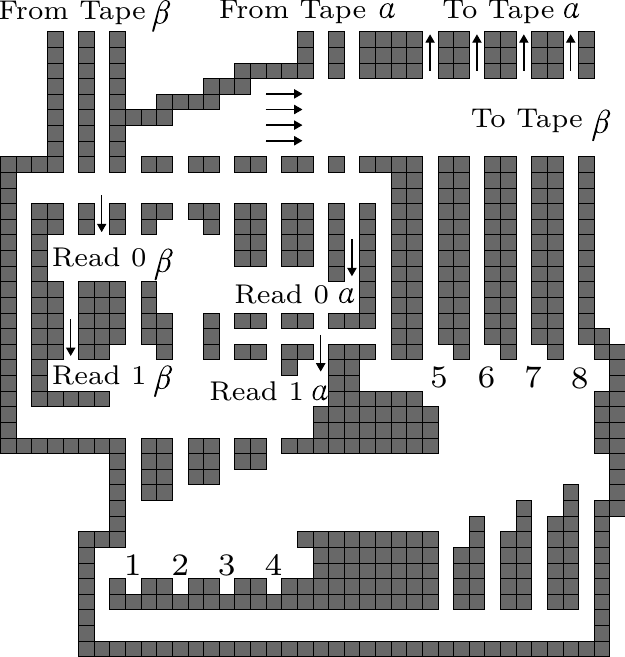}            
            \caption{State ID Gadget}\label{fig:stateid-labeled}
        \end{subfigure}
        \begin{subfigure}[b]{0.27\textwidth}\centering
            \includegraphics[height=5.9cm]{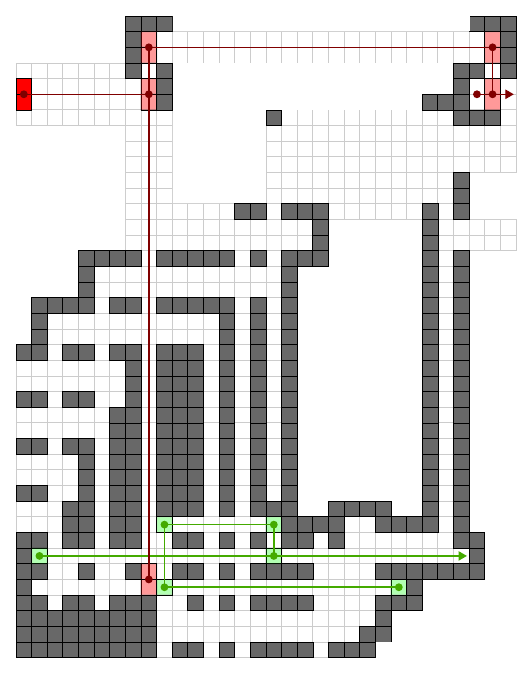}
            \caption{Reconfig. Mod.}\label{fig:ModdedTape}
        \end{subfigure}
        \caption{(a) The \emph{Repeater gadget}. Every input tunnel forwards \textsc{State$_x$ Read $y$} into the output tunnel \textsc{Forward$(x,y)$}. The \textsc{check 0} and \textsc{check 1} tunnels extend horizontally throughout the board. Every cell in the tape uses the same tunnels. (b) The \emph{State ID gadget}. If the Turing machine is in its state, it receives the repeater-dominos from each Tape's \emph{Repeater gadgets} at \textsc{From Tape}. Which output tunnel the head-dominos take at their \textsc{To Tape} encodes what the current cell on both tapes read. There is one gadget per different state.
        (c) A modification to the \emph{Tape gadget} in order to prove the complexity of reconfiguration. To set all {data-singletons} on the tape to 1, this gadget receives the \domino\ from the top left. Then, if the {data-singleton} is in \textsc{Track 0}, the \domino\ moves it to track 1. Whether the {data-singleton} was moved or not, the \domino\ then moves to the \statecell's right neighbor to repeat the process.} \label{fig:repeater}
\end{figure}

\begin{itemize}\setlength\itemsep{0em}
    \item Repeater-dominos. The orange dominos in Figure~\ref{fig:repeater}. These dominos are used to forward information from the \domino\ to the new gadgets and interact with the information from another tape.
    \item ID-domino. The blue domino in the \emph{State ID gadget} that assists with sharing the information of the two repeater-dominos, shown in Figure~\ref{fig:stateid-interactions}.
    \item Marker-dominos. A set of pink dominos in one of the \emph{Marker gadgets} on each given tape. These dominos cycle below the current \statecell\ as shown in Figure~\ref{fig:marker-cycle} and redirect the repeater-domino into the modified \emph{State gadget}. Each Tape needs a marker domino per State-move-write combination. Thus, there are $4\times|Q|$ marker dominos on each tape.
\end{itemize}

\para{Two-tape TM Gadgets.} Additional gadgets allow for the information of two different \statecell s to interact without modifying the single-tape \statecell\ too drastically. 
\begin{itemize}\setlength\itemsep{0em}
    \item \emph{State gadget} (Sec.~\ref{sssec:modified-state}). The Modified \emph{State gadget} allows the \domino\ to share the symbol read with the \domino\ from the other tape without leaving the \statecell.
    \item \emph{Repeater gadget} (Sec.~\ref{sssec:repeater}). The \emph{Repeater gadget} reads the information from the \domino\ and copies it into a {repeater-domino}. It then sends that {repeater-domino} to the correct \emph{State ID gadget}. Every \statecell\ has a \emph{Repeater gadget}.
    \item \emph{State ID gadget} (Sec.~\ref{sssec:stateid}). The \emph{State ID gadget} receives the information from both tapes and identifies which transition each tape needs to execute. There is a \emph{State ID gadget} for every state.
    \item \emph{Marker gadget} (Sec.~\ref{sssec:marker}). The \emph{Marker gadget} marks which \statecell\ the \domino\ is at and redirects the transition from the \emph{State ID gadget} into that \statecell's \emph{Modified State Update}. Every \statecell\ has a \emph{Marker gadget}.
    \item \emph{Repeater Reset gadget} (Sec.~\ref{sssec:repeater-reset}). The \emph{Repeater Reset gadget} catches the {repeater-domino} and resets it back to its initial tunnel. There is one \emph{Repeater Reset gadget} per tape.
    \item \emph{Modified State Update} (Sec.~\ref{sssec:state-update-multitape}). The \emph{Modified State Update} is the section of the  Modified \emph{State Gadget} responsible for updating the state according to the information read. The process is similar to the one in Figure~\ref{fig:stateup}.
\end{itemize}

A diagram of a two-tape single-cell Turing machine with the attached two-tape gadgets is shown in Figure \ref{fig:multitape-diagram}. Following the arrows, we can see how the machine works. First, each \statecell\ reads its symbol and its state. Then, the \emph{Repeater gadget} forwards the symbol from each \statecell\ to the \emph{State ID gadget} that matches the state. In the \emph{State ID gadget}, the two repeated signals merge their information to know what specific transition the Turing machine should execute. This transition is then sent to the \emph{Marker gadget} through a tunnel unique to each transition. Then, the Turing machine executes the transition in the cell, the same as in Section \ref{sec:singletape}.

\begin{figure}[t]
    \centering
    \includegraphics[width=.9\linewidth]{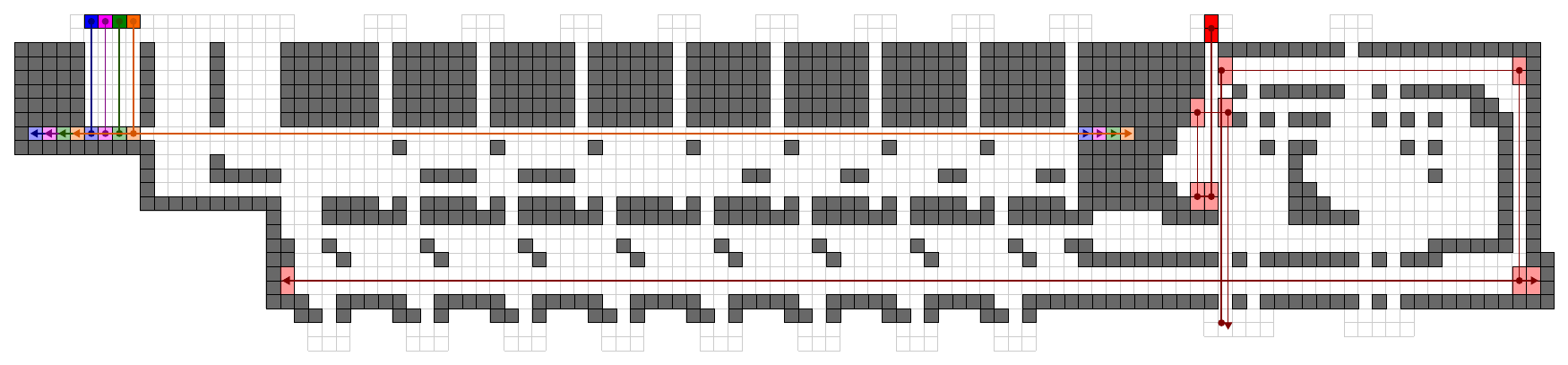}
    \caption{Modified \emph{State gadget}. The \domino\  reads the 
    state and goes down to the \emph{Repeater gadget}, and then returns and waits in the bottom tunnel. The {State-singletons} cycle at \textsc{Wait}.}
    \label{fig:multitape-state}
\end{figure}

\subsection{Modified State Gadget}\label{sssec:modified-state}

We reuse the \emph{State gadget} from \ref{sssec:state-gadget}, with a modification that sends the \domino\ down into the \emph{Repeater gadget}. Then, the modified \emph{State gadget} catches the \domino\ on the next up tilt and sends it into the left half of the modified \emph{State gadget}, where it waits for the transition to update the state as described in Section \ref{sssec:state-update-multitape}. This is shown in Figure \ref{fig:multitape-state}.

\subsection{Repeater Gadget}\label{sssec:repeater}
\begin{figure}[t]
        \centering
        \begin{subfigure}[b]{0.32\textwidth}\centering
            \includegraphics[width=.9\textwidth]{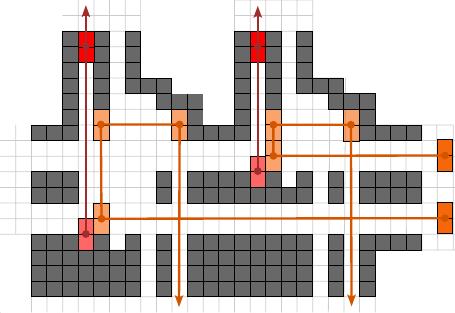}
            \caption{}\label{fig:repeater-state0}
        \end{subfigure}
        \begin{subfigure}[b]{0.32\textwidth}\centering
            \includegraphics[width=.9\textwidth]{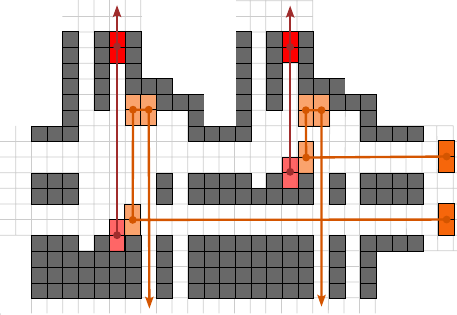}            
            \caption{}\label{fig:repeater-state1}
        \end{subfigure}
        \begin{subfigure}[b]{0.32\textwidth}\centering
            \includegraphics[width=1.\textwidth]{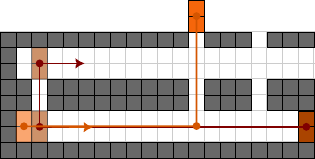}
            \caption{}\label{fig:repeater-reset}
        \end{subfigure}
        \caption{(a) The \emph{Repeater gadget} forwarding a symbol in state $q_1$.(b) The \emph{Repeater gadget} forwarding a symbol in state $q_0$. (c) The \emph{Repeater Reset gadget}. This gadget is below the leftmost \emph{Marker gadget}.}
        \label{fig:repeater-interacitons}
\end{figure}

After the modified \emph{State gadget} sends the \domino\ out of the \statecell, the information carried by the \domino\ needs to interact with the information from the other tape. We use the \emph{Repeater gadget} shown in Figure \ref{fig:repeater-labeled} to copy this information into a {repeater-domino} and send each \domino\ back into its own \statecell. This gadget has two orange vertical dominos traversing freely left through \textsc{check 0} and \textsc{check 1} as shown in Figure \ref{fig:repeater-state0} and \ref{fig:repeater-state1}. The gadget receives the \domino\ from the modified \emph{State gadget} in the cell on a down tilt. On the following left tilt, the \domino\ blocks one of the two \textsc{check} tunnels. The tunnel blocked indicates the value read, and the position represents the machine's current state. The different interactions of the \domino\ and the orange {repeater-dominos} are shown in Figure \ref{fig:repeater-state0} and \ref{fig:repeater-state1}. After the interaction, the \domino\ goes back up into the \statecell\ and moves into the \emph{State gadget}. One {repeater-domino} exits the \emph{Repeater gadget} through one of the four tunnels in its bottom. There is a unique tunnel for each of the four possible interactions. These tunnels send the {repeater-domino} to the corresponding \emph{State ID gadget}. Each \statecell\ has its own \emph{Repeater gadget}. A \emph{Repeater gadget} from Tape $\alpha$ is in line with the other \emph{Repeater gadgets} from all \statecell\ in Tape $\alpha$.

\subsection{State ID Gadget}\label{sssec:stateid} 

\begin{figure}[t]
        \centering
        \begin{subfigure}[b]{0.44\textwidth}\centering
            \includegraphics[width=.8\textwidth]{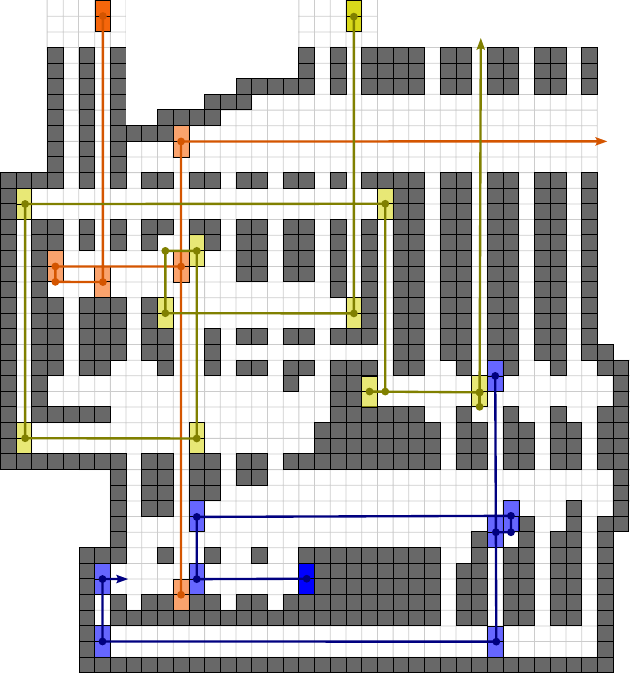}
            \caption{$\alpha$ read 0, $\beta$ read 0}\label{fig:SID00}
        \end{subfigure}
        \begin{subfigure}[b]{0.44\textwidth}\centering
            \includegraphics[width=.8\textwidth]{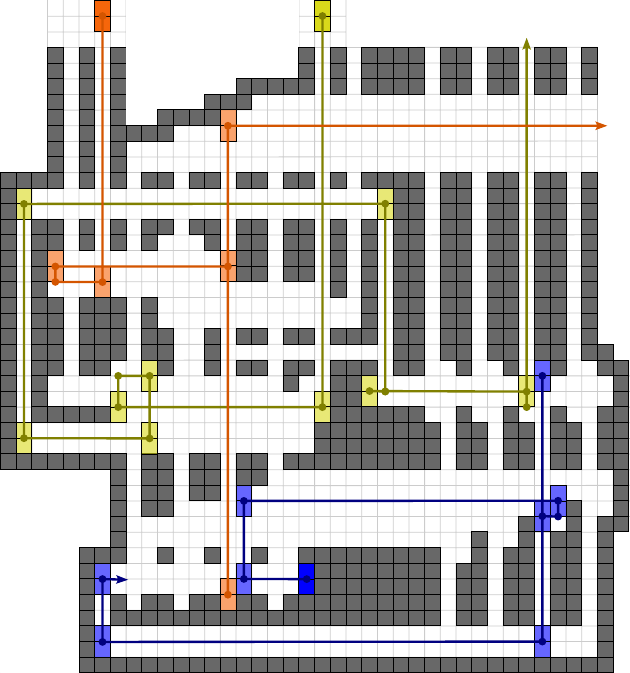}
            \caption{$\alpha$ read 1, $\beta$ read 0}\label{fig:SID01}
        \end{subfigure}
        \begin{subfigure}[b]{0.44\textwidth}\centering
            \includegraphics[width=.8\textwidth]{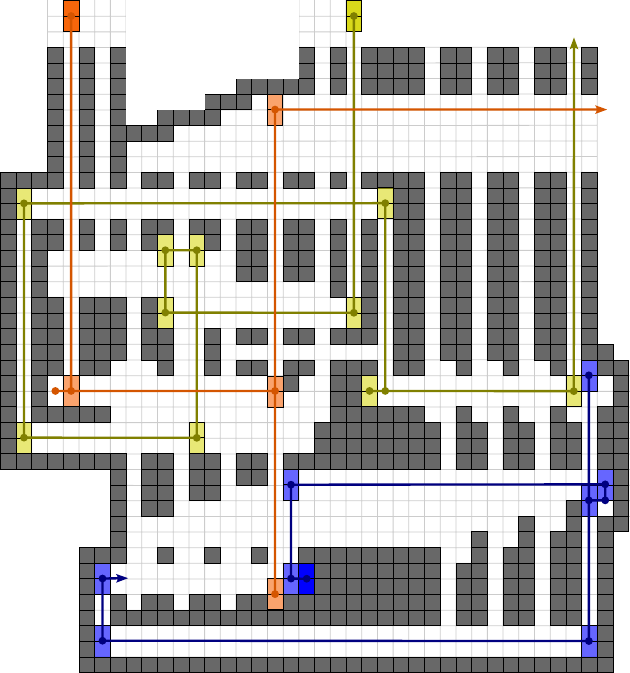}
            \caption{$\alpha$ read 0, $\beta$ read 1}\label{fig:SID10}
        \end{subfigure}
        \begin{subfigure}[b]{0.44\textwidth}\centering
            \includegraphics[width=.8\textwidth]{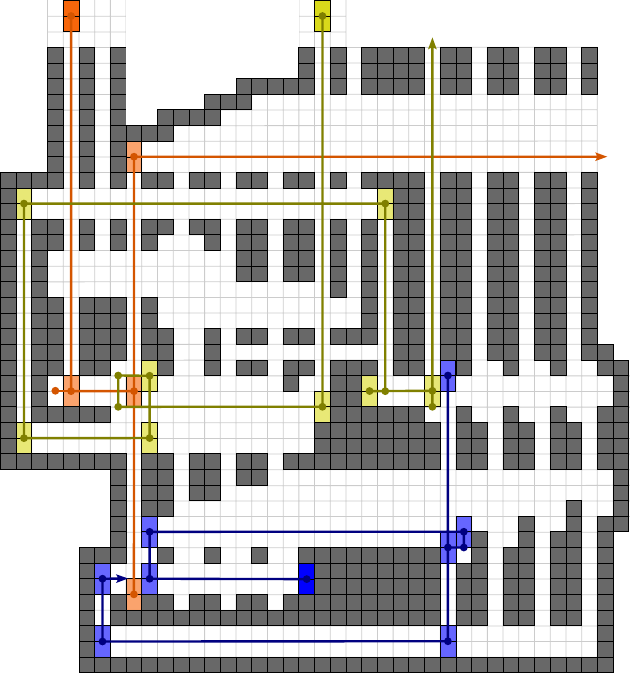}
            \caption{$\alpha$ read 1, $\beta$ read 1}\label{fig:SID11}
        \end{subfigure}
        \caption{The four possible interactions in a \emph{State ID gadget}. Depending on the symbols read by both tapes, different output tunnels will be used by the yellow (Tape $\alpha$) and orange (Tape $\beta$) repeater-dominos.}
        \label{fig:stateid-interactions}
\end{figure}

This gadget takes as input the repeater-domino. Since there is a \emph{Repeater gadget} for each cell, there will always be two repeater-dominos coming into the \emph{State ID gadget}, one for Tape $\alpha$ and one for Tape $\beta$, as shown in Figure \ref{fig:stateid-interactions}. The purpose of the \emph{State ID gadget} is to read the symbols from both tapes and send the same {repeater-domino} to carry that information back to \statecell. Each of the four different combinations of symbols between the two tapes has a tunnel to carry this information, as shown in Figure \ref{fig:stateid-labeled}. We use the yellow domino as the repeater-domino from Tape $\alpha$ and the orange domino as the repeater-domino from Tape $\beta$. The two dominos enter the gadget in a position denoting the tape they come from and the symbol they read. According to their position, they can interact in four different ways. According to the interaction, the orange domino will fall into one of the positions labeled 1-4 in Figure~\ref{fig:stateid-labeled}. This will stop the blue ID-domino so it can stop the yellow domino in the correct position of those labeled 5-8. This way, the orange and yellow dominos are each positioned according to the two symbols, and can separately take the correct transition to their tape. The ID-domino then makes its way back to its starting position. The four different interactions are shown in Figure \ref{fig:stateid-interactions}.

\subsection{Marker Gadget}\label{sssec:marker}

\begin{figure}[t]
    \centering
    \begin{subfigure}[b]{0.47\textwidth}\centering
        \includegraphics[width=1.\linewidth]{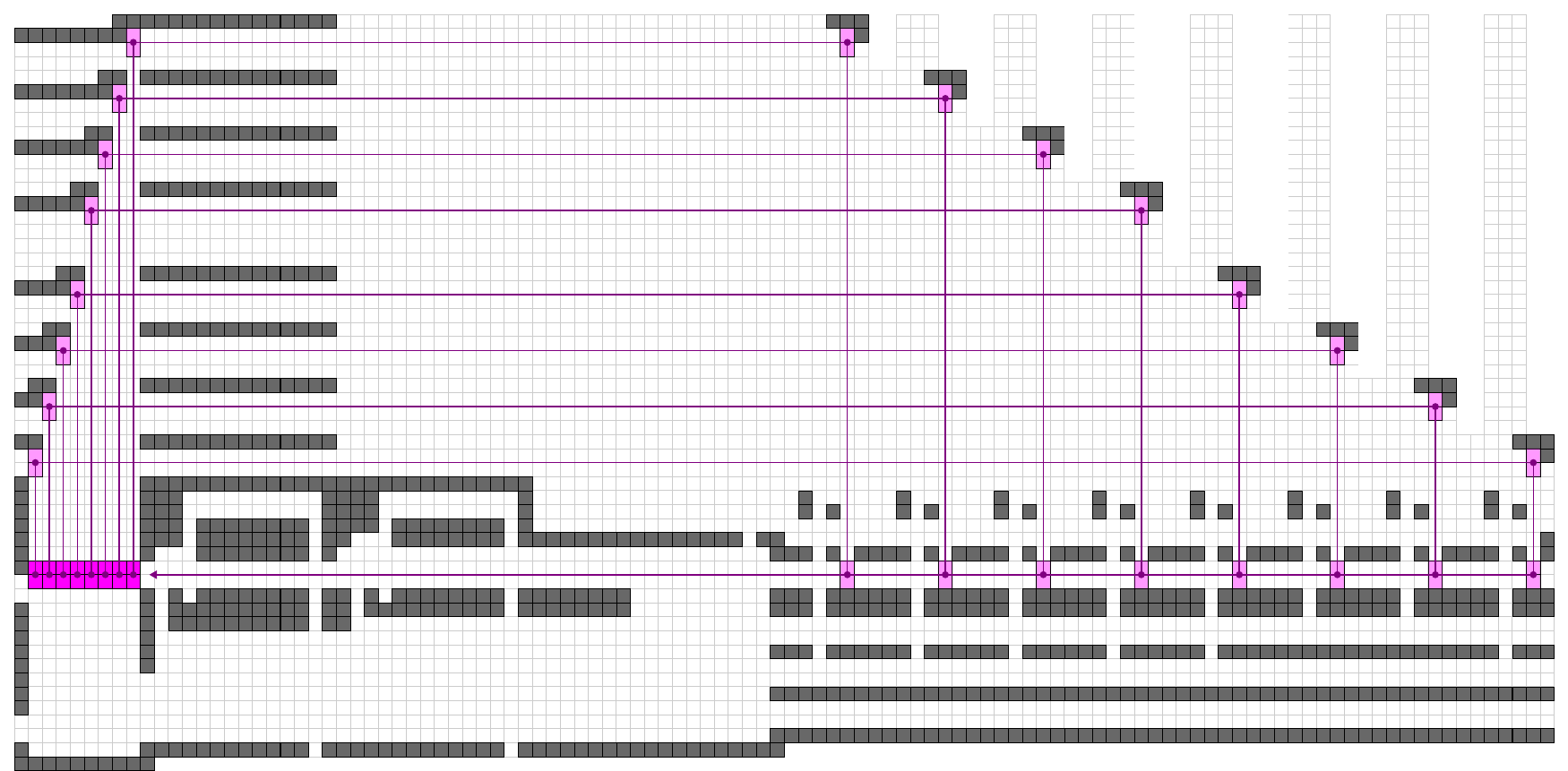}
        \caption{}
        \label{fig:marker-cycle}
    \end{subfigure}
    \begin{subfigure}[b]{0.47\textwidth}\centering
        \includegraphics[width=1.\linewidth]{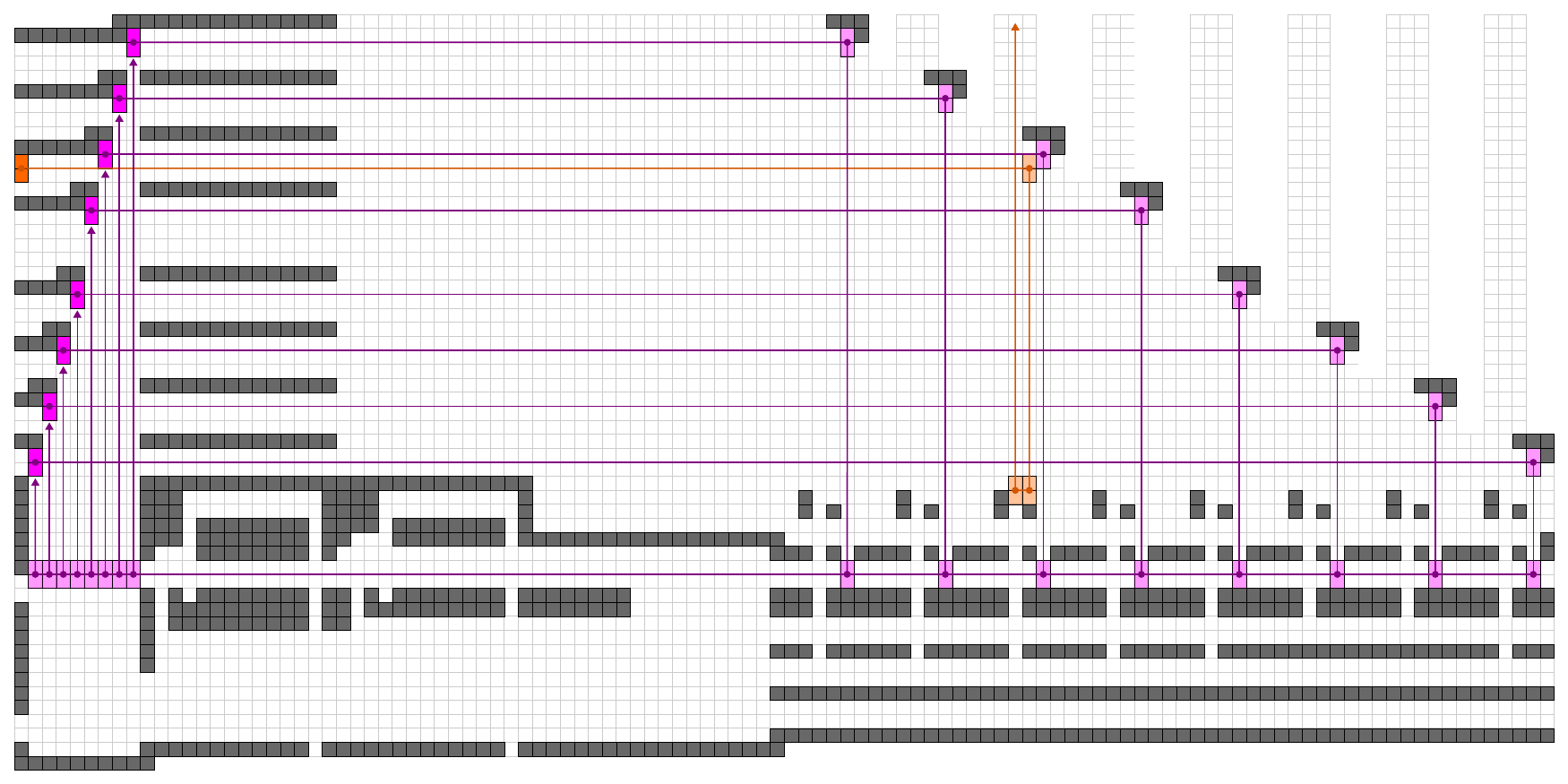}
        \caption{}
        \label{fig:marker-up}
    \end{subfigure}
    \begin{subfigure}[b]{0.47\textwidth}\centering
            \includegraphics[width=1.\textwidth]{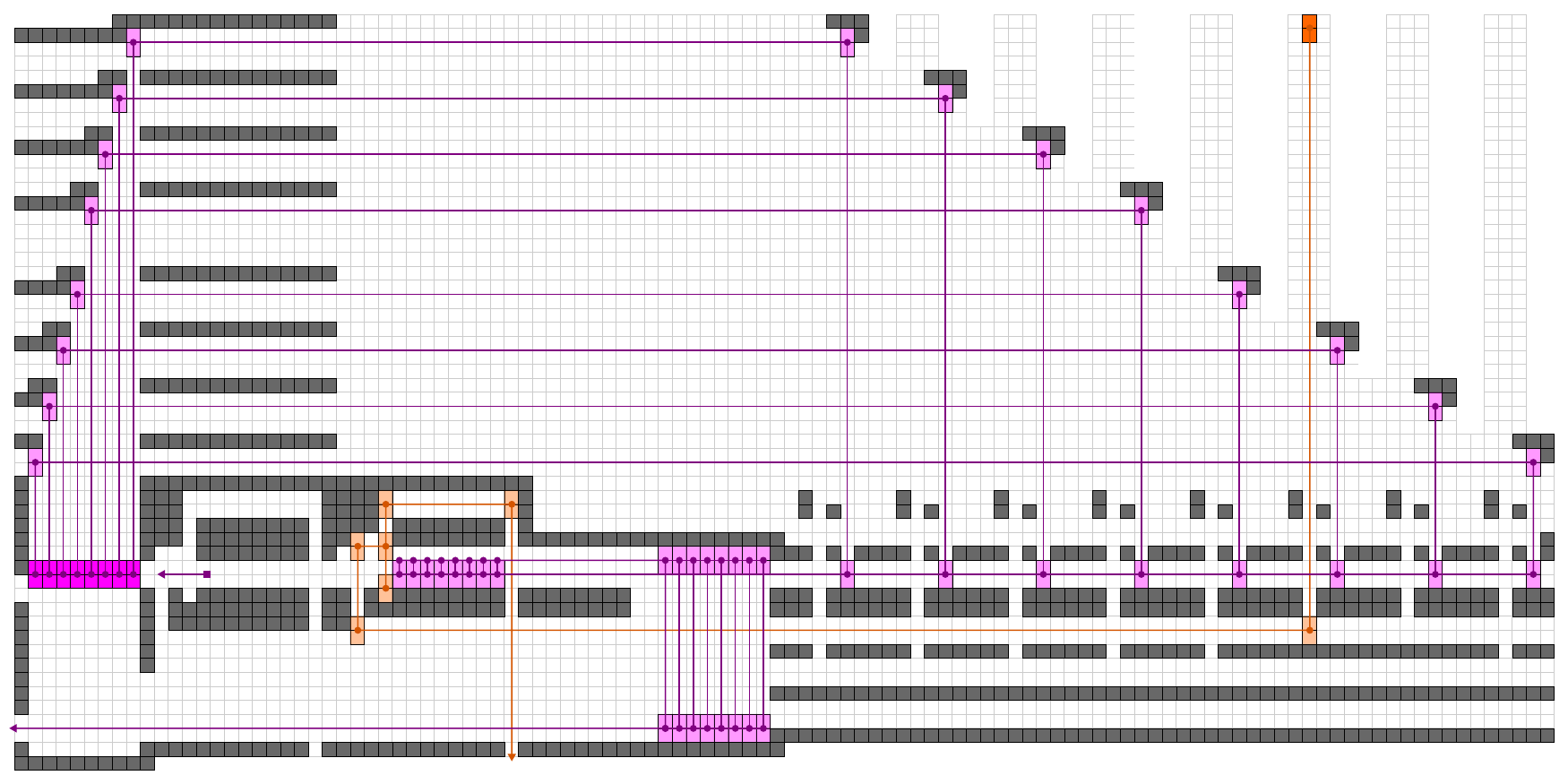}
            \caption{}\label{fig:marker-send-left}
        \end{subfigure}
        \begin{subfigure}[b]{0.47\textwidth}\centering
            \includegraphics[width=1.\textwidth]{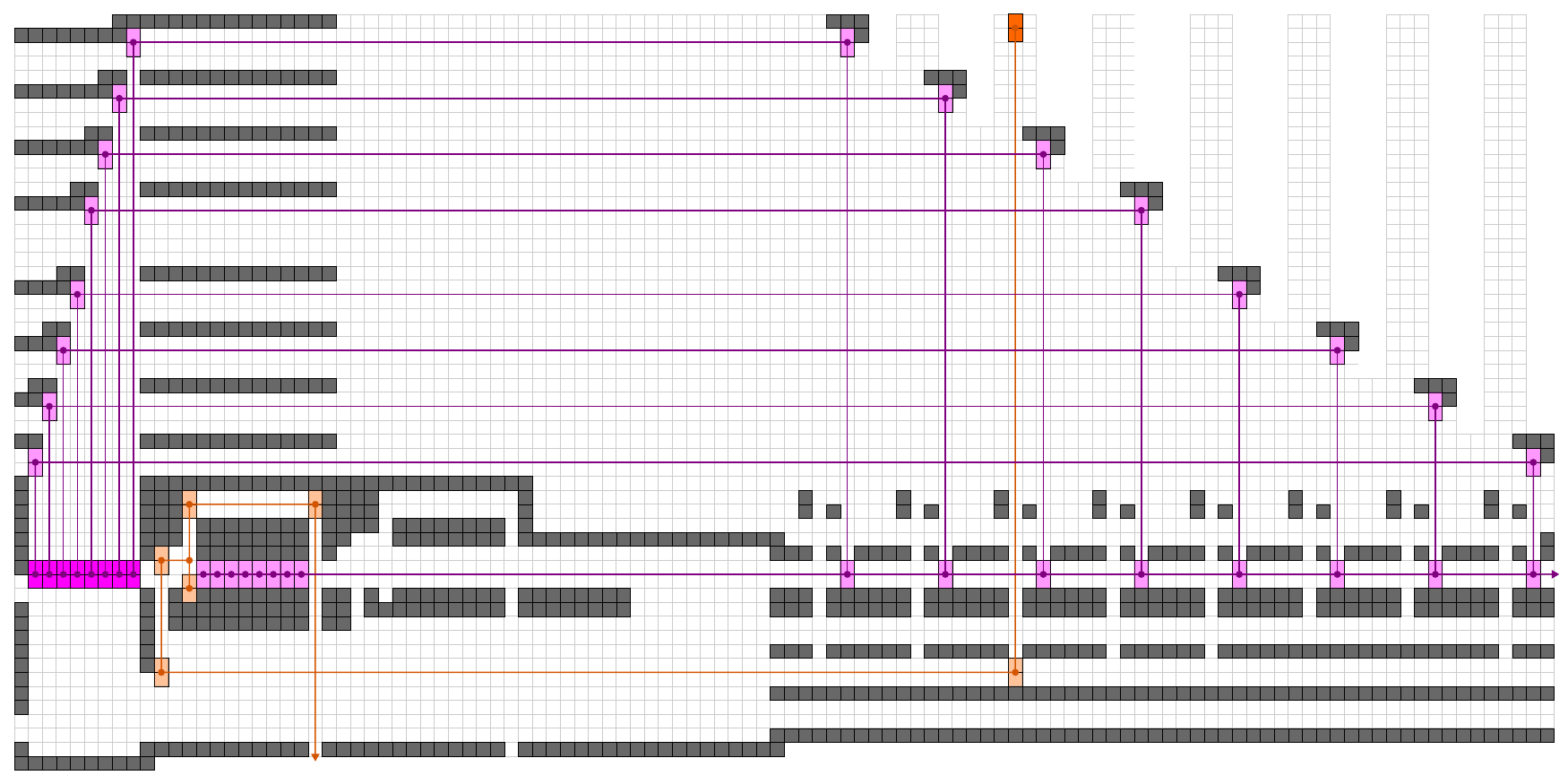}
            \caption{}\label{fig:marker-send-right}
        \end{subfigure}
    \caption{(a) The marker-dominos cycling in the \emph{Marker gadget}. (b) The repeater-domino being redirected up into the modified \emph{State gadget}. (c) The repeater-domino sending the marker-dominos to the right and leaving to the \emph{Repeater Reset gadget}. (d) The repeater-domino sending the marker-dominos to the left and leaving to the \emph{Repeater Reset gadget}. }\label{fig:markerdomsall}
\end{figure}

The \emph{Marker gadget} keeps track of which cell the \domino\ is currently reading and allows the {repeater-dominos} leaving the \emph{State ID gadget} to identify which \statecell\ construct they should stop at. The \emph{Marker gadget} uses pink marker-dominos that cycle as shown in Figure \ref{fig:marker-cycle}. There are $4 \times |Q|$ marker-dominos per tape on the board at all times. When the repeater-domino enters (after a right tilt), one of the marker-dominos will catch the repeater-domino and send it up into the modified \emph{State gadget}, as shown in Figure \ref{fig:marker-up}. Then, the same repeater-domino is redirected back down into one of the two tunnels labeled `Send Left Indicator' and `Send Right Indicator' in the bottom part of the \emph{Marker gadget}. Depending on which tunnel the repeater-domino stops at, it will send the marker-dominos right or left to the \emph{Marker gadget} of the neighboring \statecell, as shown in Figures \ref{fig:marker-send-left} and \ref{fig:marker-send-right}. An alternating height difference is needed for this gadget due to the sending and receiving mechanic, similar to the one in the \statecell\ described in Section \ref{ssec:alternating}.


\begin{figure}[t]
    \centering
    \includegraphics[width=1.\linewidth]{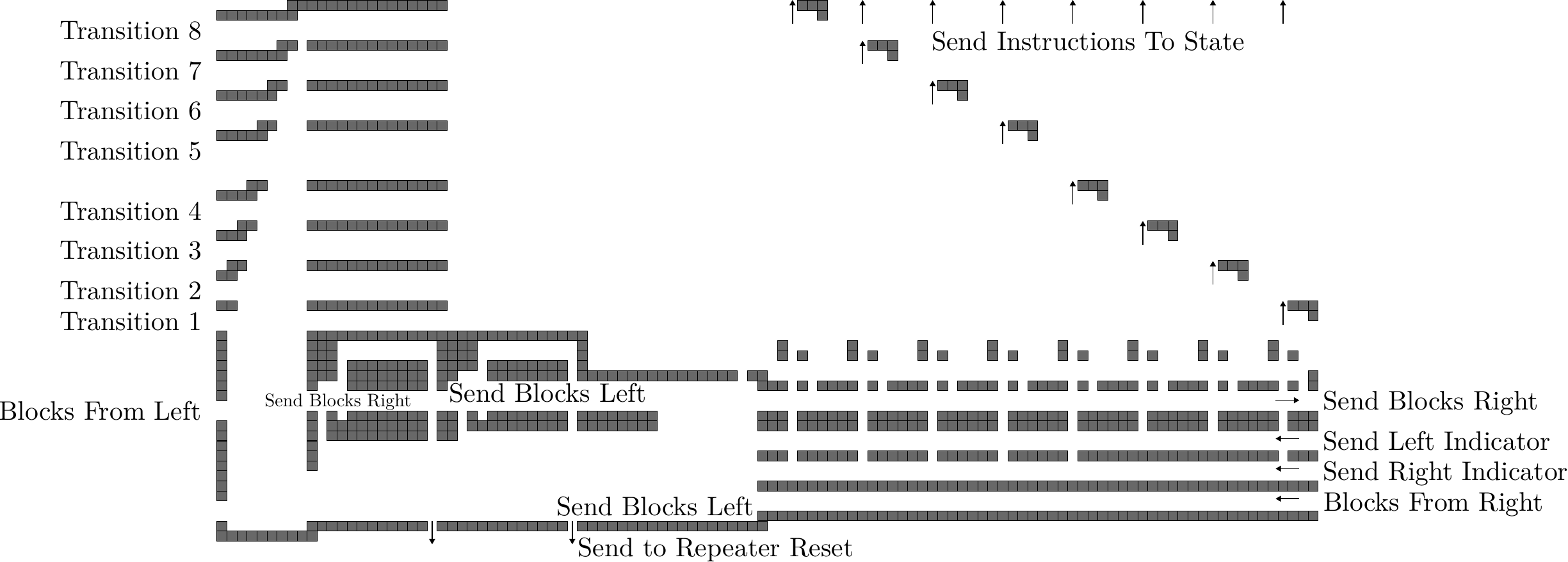}
    \caption{\emph{Marker gadget}. The top half of the gadget is where the repeater-domino gets stopped and redirected to the \emph{state gadget} to execute the transition output its location indicates. In the bottom half, the repeater-domino sends the marker-dominos left or right, depending on the transition's direction.} \label{fig:marker-labeled}
\end{figure}

\subsection{Repeater Reset Gadget}\label{sssec:repeater-reset}
Once out of the \emph{Marker gadget}, the repeater-domino needs to return to its original tunnel in the \emph{Repeater gadget}. To do this, it gets dropped into the tunnel labeled `check for 1' in Figure \ref{fig:repeater-labeled}. Then, in the case that there already is a domino in the tunnel, one of the two dominos will be pushed up into the `check for 0' tunnel. This interaction is shown in Figure \ref{fig:repeater-reset}. Note that the \textsc{check 1} and \textsc{check 0} tunnels extend horizontally from the \emph{Repeater Reset gadget} all the way to the rightmost \emph{Repeater gadget}.

\begin{figure}[t]
    \centering
    \includegraphics[width=.9\linewidth]{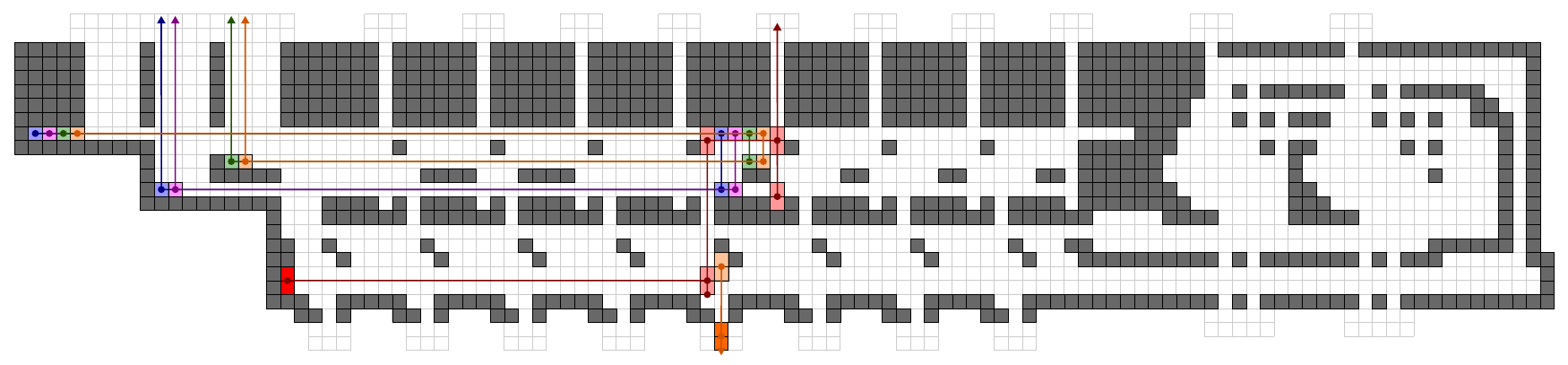}
    \caption{The \domino\ executing a state update transition. The repeater-domino blocks the \domino, sending it up. On the following right tilt the \domino\ stops the singletons. Then, the process of separating active-singletons and inert-singletons, and the rest of the information processing in the cell, follows the same process as in Section \ref{sssec:state-gadget}.}
    \label{fig:state-updater}
\end{figure}

\subsection{Modified State Update}\label{sssec:state-update-multitape}
The way the state is updated is similar to the one in the \emph{State gadget} in Section \ref{sssec:state-gadget}. The \domino\ comes in from the \emph{State gadget} and is trapped in the bottom part tunnel. When the repeater-domino from the \emph{Marker gadget} comes up, it will stop the \domino\ and send it up to execute a transition to change state, as shown in Figure \ref{fig:state-updater}. The \domino\ stops the {state-singletons} such that upon tilt down, they are separated into {inert-singletons} and {active-singletons}. Then, the singletons and the \domino\ continue to their respective gadgets described in Section \ref{sec:singletape}.
Using these gadgets, along with the \statecell\ from Section~\ref{sec:singletape}, we can establish the following:

\begin{theorem}\label{thm:2tape}
    For any two-tape Turing machine $\mathcal{M}_2$ with $s=|Q|$ states and a space bound $n$ such that each tape of $\mathcal{M}_2$ uses at most $n$ tape cells, there exists a non-bonding rotational Full-Tilt simulation of the machine with board size $O(ns^3)$ that simulates the machine at a rate of one step per $O(1)$ rotations.
\end{theorem}
Given a two-tape Turing machine $\mathcal{M}_2$, we design the board as shown in Figure~\ref{fig:multitape-diagram}. There are $2n$ \statecell s with height offset to separate Tape $\alpha$ from Tape $\beta$, $2n$ \emph{Marker gadgets} below the \statecell s, and $s$ \emph{State ID gadgets} in the bottom left corner with a height and width offset to separate the input and output tunnels. A single \emph{Repeater Reset gadget} is placed below the leftmost \emph{Marker gadget} and is connected to the checker-tunnels from the \emph{Repeater gadgets}. For every additional state, the \emph{Repeater}, \emph{State ID}, and \emph{Marker gadgets} need an extra pair of input/output tunnels. This construction will require $O(s)$ domino particles. This yields a board of height $O(s)$ and width $O(ns^2)$.

\subsection{The Single-Step Model}
Similar to the single-tape modification in Section \ref{ssec:sstm}, we modify the two-tape machine for the Single-Step model. As before, the \statecell\ spacing is bounded, but 
we can not achieve a constant number of cycles for each step of the TM. Since the two tape heads need to communicate, the distance between them may be as far as the length of the tape ($T$). Thus, the number of cycles per step is $O(T)$ for any constant cycle.

\begin{corollary}\label{thm:SS2tape}
    For any two-tape Turing machine $\mathcal{M}_2$ with $s=|Q|$ states and a space bound $n$ such that each tape of $\mathcal{M}_2$ uses at most $n$ tape cells, there exists a non-bonding rotational Single-Step simulation of the machine with board size $O(ns^3)$ that simulates the machine at a rate of one step per $O(T)$ rotations.
\end{corollary}

\para{Systolic Arrays.}
In some sense, the Single-Step model's limited travel range prevents many possible efficiency improvements over the single-tape TM. However, due to the use of the \statecell\ constructions, multiple computations could be occurring simultaneously in different parts of the tape if multiple dominos and state-singletons are included. This would require coordination to prevent collisions and limit each computation to a section of the tape. 

We can easily extend the single-tape TM construction to work as a linear systolic array \cite{Brent:1984:ToC,Kung:1982:C,Mead:1980:BOOK} with a single \domino\ at each head location and a set of state-singletons. The tape is split into $k$ independent memory sections, each with its own tape head (\domino\ and state-singletons). The left/right boundary \statecell s of each memory section has a dual-rail \emph{input} area on the left cell and an \emph{output} area on the right cell for the \domino\ to pass on information to the neighboring memory areas.

\begin{corollary}\label{thm:SS2systolic}
    For any single-tape Turing machine $\mathcal{M}$ with $s=|Q|$ states, $k$ heads (each with $n/k$ tape cells), and a tape of length $n$, there exists a non-bonding rotational Single-Step simulation of the machine with board size $O(ns^3)$ that simulates the machine at a rate of one step of each head per $O(1)$ rotations.
\end{corollary}

The Full-Tilt model can also achieve this, but since it can also implement a two-tape TM without slowdown, the result is trivial.
\section{Efficient Circuit Simulation}\label{sec:tc}

In this section, we show how to efficiently simulate Threshold circuits of any level.
The results in \cite{Becker2019} show that we can implement computationally universal circuits with polynomial board constructions, even for unbounded \textsc{FAN-IN}, with logic gates that compute the results in a constant number of $\langle d, l/r, d, l/r \rangle$ cycles. We provide similar constructions that solve decision problems using gates designed for a fixed $\langle u, r, d, l \rangle$ cycle, and then extend the work with a \textsc{Majority}
gate to allow for efficient simulation of \textit{Threshold Circuits}. We improve upon the limitations of a digital sorting network, and show that we can produce monotonic permutations of Boolean values in a constant number of $\langle u, r, d, l \rangle$ rotations. By extension, we give an N-\textsc{Majority} gate gadget in $O(1)$ rotations and $O(n^2)$ size for unbounded, unweighted \textsc{FAN-IN} with $n$ inputs, and show that a Threshold Circuit of depth $d$ can be simulated in a polynomial board in $O(d)$ cycles. 

\begin{figure}[t]  
    \centering
	\begin{subfigure}[b]{0.32\textwidth}\centering
        \includegraphics[width=1.\textwidth]{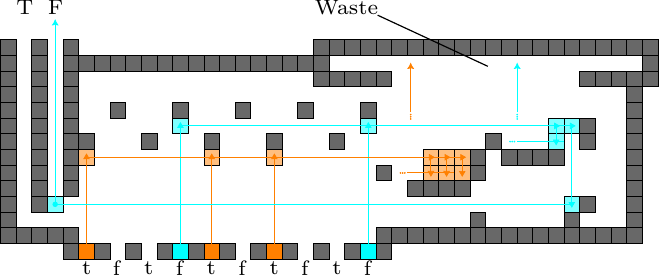}
        \caption{\textsc{AND Gate}} \label{fig:AND gate}
    \end{subfigure}
	\begin{subfigure}[b]{0.32\textwidth}\centering
        \includegraphics[width=1.\textwidth]{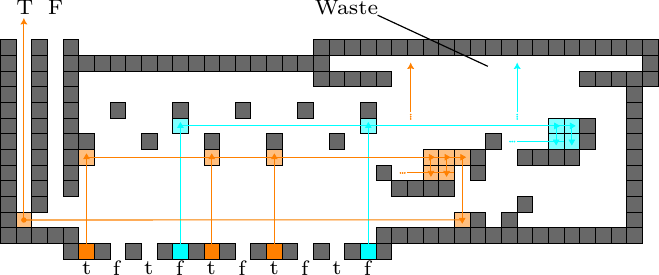}
        \caption{\textsc{OR Gate}} \label{fig:OR gate}
    \end{subfigure}
    \begin{subfigure}[b]{0.32\textwidth}\centering 
     \includegraphics[width=1.\textwidth]{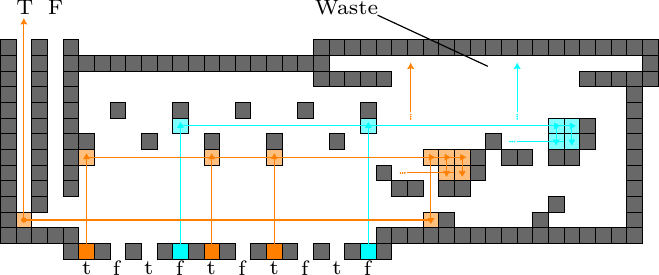}
        \caption{\textsc{Majority Gate}} \label{fig:Majority gate}
    \end{subfigure} 
    \caption{(a) 5-\textsc{AND} gate with 3-$True$, 2-$False$ inputs. (b) 5-\textsc{OR} gate with 3-$True$, 2-$False$ inputs. Each gate actually computes $\textit{two}$ expressions that can only be $True$ when the other is $False$. Both gates compute their $True$ and $False$ expressions in parallel in such a way that the output is always either $True$ or $False$ in dual-rail. Notice that you can take the Boolean inversion of an operand by swapping its respective filters.
    (c) A dual-rail 5-\textsc{Majority} gate with inputs $\langle 1, 0, 1, 1, 0 \rangle$. Note that, like the Turing Machine, these gates operate with repeated executions of $\langle u, r, d, l \rangle$ cycles.}
    \label{fig:logic-gates}
\end{figure}


Being topologically atomic in their design (in terms of computation time), similar to the implementations in \cite{Becker2019}, the gates simulate Boolean circuits. However, a limitation of conventional digital logic gates is their inability to efficiently compute \textit{threshold} gates. Consider a \textsc{Majority} gate, where the expression is $True$ when $\lceil\text{FAN-IN}/2\rceil$ operands are $True$, and $False$ otherwise, for odd-parity length inputs.\footnote{Even-parity lengths are often implementation-defined and thus not addressed.}

When the \textit{quantity} of $True$ operands is known for a given gate, determining if the quantity is greater than $k$ is trivial, i.e., the $k^{th}$ value of a monotonic sequence of operands $T$. Then $T[k]$ is the \textsc{Majority}. A sorting network permutes $T$ as 
$\langle 1, 0, 1, 1, 0, \dots \rangle \mapsto \langle 0, 0, \dots, 1, 1, 1 \rangle$.




\begin{figure}[t]
    \centering
	\begin{subfigure}[b]{0.41\textwidth}\centering
        \includegraphics[height=3cm]{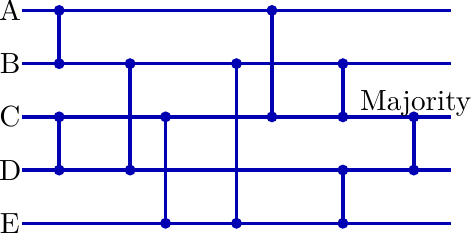}
        \caption{\textsc{5-Sorting-Network} with  \textsc{Majority}}
    \end{subfigure}
	\begin{subfigure}[b]{0.49\textwidth}\centering
        \includegraphics[height=3cm]{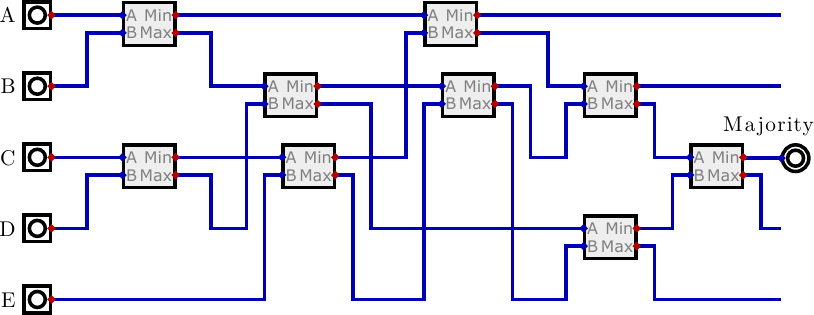}
        \caption{Digital circuit implementation of a \textsc{Majority}}
    \end{subfigure}
    \caption{
    A 5-\textsc{Majority} gate using a sorting network, where the $3^{rd}$ operand indicates the \textsc{Majority}.
    }
    \label{fig:5-MAJORITY implementation}
\end{figure}

An intrinsic property of sorting networks is that comparisons between channels are performed in parallel, thus greatly minimizing the depth requirements of comparing and swapping operands for an optimal sorting network (Figure \ref{fig:5-MAJORITY implementation}). Numerous algorithms and implementations exist for producing optimal or near-optimal sorting networks with depth $O(\log^2 n)$ and size $O(n \log^2 n)$ where $n$ is the number of inputs (assume comparators are $O(1)$), such as MergeSort and the \textsc{Pairwise Sorting Network} \cite{Parberry:1992}. A limitation of sorting networks is that the optimality of a network of arbitrary input size is unknown, and determining whether a configuration of a network is a valid sorting network is coNP-complete \cite{Parberry:1990}. Both of these factors make a \textsc{Sorting-Network Majority} gate sub-optimal for sufficient input sizes in Full-Tilt. Thus, we provide an alternative implementation that improves upon digital Boolean sorting within Full-Tilt. 

\begin{figure}[t] 
    \centering
	\begin{subfigure}[b]{0.5\textwidth}\centering
	   \includegraphics[width=1.\textwidth]{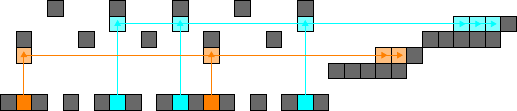}
       \caption{Sorting}\label{fig:boolean-sorting-gadget}    
    \end{subfigure}
    \begin{subfigure}[b]{0.4\textwidth}\centering
        \includegraphics[width=.9\textwidth]{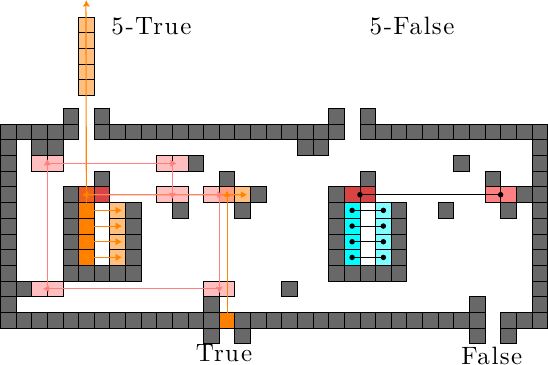}
        \caption{5-\textsc{FAN-OUT}}\label{fig:5-FAN-OUT}    
    \end{subfigure}
    \caption{(a) A simple filter gadget that groups $True$ and $False$ dual-rail operands (represented as orange tiles (left) and cyan tiles (right), respectively) into monotonic groups of $True$ and $False$ tiles, each in their own row. This gadget demonstrates sorting of an arbitrary number of Boolean operands within a single $\langle u, r, d, l \rangle$ cycle (specifically achieving the monotonic grouping in $\langle u, r \rangle$) with $O(n)$ space for $n$ input operands. This grouping is how $\text{AND}$, $\text{OR}$, and \textsc{Majority} gates are simulated with $\langle u, r, d, l \rangle$ cycles.
    (b) A 5-FAN-OUT gate gadget that outputs 5 tiles from the $True$ side of a dual-rail Boolean gadget when the input is $True$, and from the $False$ side when the input is $False$. Dominos (red) block the FAN-OUT tiles until they are unlatched by the input. When a domino is unlatched, 5 tiles exit from the 5-FAN-OUT gate at the beginning of the second $\langle u, r, d, l \rangle$ cycle, and the unlatched domino will re-latch at the end of the cycle. The output tiles then get separated into the corresponding 5 output dual-rail paths (the reverse of sorting).
    }
\end{figure}

\para{\textsc{FAN-OUT}.} We note that \cite{Becker2019} proved that in this model, a \textsc{FAN-OUT} requires a domino within the gadget. In order to duplicate the boolean value of some output, additional tiles must be stored, and a domino is required to keep the tiles trapped until the \textsc{FAN-OUT} is activated. Thus, in order to maintain efficiency, a domino exists at every \textsc{FAN-OUT} gate. We give a cleaner arbitrary \textsc{FAN-OUT} in Figure \ref{fig:5-FAN-OUT} that uses two dominos. It has the possible output tiles trapped internally, which requires a height of $O(n)$ where $n$ is the size of the \textsc{FAN-OUT} output. This affects the height of the board, so we note that the \textsc{FAN-OUT} in \cite{Becker2019} has constant height, even if it requires sending the output tiles in as input. Thus, we use the constant height of all gadgets in our construction to achieve the following.

\begin{theorem}\label{thm:threshold}
    For any Threshold Circuit $T$ of depth $d$ and width $w$, there exists a Full-Tilt system with a board of size $O(w^2d)$ that simulates $T$ with $O(d)$ $\langle u, r, d, l \rangle$ cycles.
\end{theorem}

\begin{proof}

An $n$-\textsc{Majority} gate computed via boolean sorting can be simulated with a Full-Tilt system with a board of size $O(n)$ that determines the \textsc{Majority} in $O(1)$ $\langle u, r, d, l \rangle$ cycles. This is done via sorting the $True$ and $False$ dual-rail input tiles into two columns (Figure~\ref{fig:boolean-sorting-gadget}), and allowing a single tile from the column with more tiles (more than $\lceil n/2 \rceil$) to exit the correct dual-rail pathway. All other tiles are trapped. We provide an example construction of a \textsc{5-Majority} gate in Figure~\ref{fig:Majority gate}.
Any gate's input value $n$ must be less than the width of the circuit $w$. The \textsc{Majority}, \textsc{AND}, \textsc{OR}, and \textsc{FAN-OUT} gates all require $O(n)$ space in width and a constant height (Figure \ref{fig:logic-gates}). Thus, the total size of the board with each gate gadget is $O(nwd)$, which is $O(w^2d)$.
\end{proof}

\begin{figure}[t]
    \centering
    \begin{subfigure}[b]{0.27\textwidth}\centering
        \includegraphics[height=4.7cm]{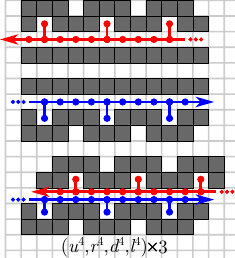}
        \caption{Single-Step Path}\label{fig:SSpath}
    \end{subfigure}
    \hspace{.2cm}
    \begin{subfigure}[b]{0.29\textwidth}\centering
        \includegraphics[height=4.7cm]{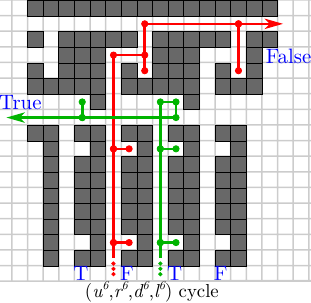}
        \caption{Single-Step Sort}\label{fig:SSsort}
    \end{subfigure}
    \hspace{.2cm}
    \begin{subfigure}[b]{0.32\textwidth}\centering
        \includegraphics[height=3.5cm]{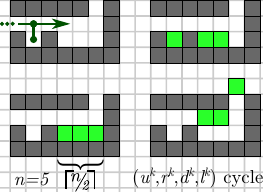}
        \caption{Single-Step Majority}\label{fig:SSmaj}
    \end{subfigure}
    \caption{
    (a) A left, right, and dual tunnel in Single-Step that is traversable for the cycle $\langle u^4,$ $r^4,$ $d^4,$ $l^4 \rangle$.
    (b) Binary sorting dual-rail input in Single-Step for two variables. All $True$ tiles exit on the same path to the left, and all $False$ variables exit the same tunnel to the right using a $\langle u^6,$ $r^6,$ $d^6,$ $l^6 \rangle$ cycle.
    (c) A Single-Step \textsc{Majority} example for some constant $4k$-length cycle with five inputs. 
    }
    \label{fig:ModdedTape2}
\end{figure}

\para{The Single-Step Model.} 
Although the basic idea is the same as with the Full-Tilt model, we incur the distance divided into a number of cycles based on the distance moved per cycle. This is demonstrated in Figure \ref{fig:SSpath}. The gates must be similarly modified, but can be done so with a relatively small cycle length.

\begin{corollary}\label{thm:SSthreshold}
    For any Threshold Circuit $T$ of depth $d$ and width $w$, there exists a Single-Step system with a board of size $O(w^2d)$ that simulates $T$ with $O(dw)$ $\langle u, r, d, l \rangle$ cycles.
\end{corollary}

\begin{proof}
Similar to Theorem \ref{thm:threshold}, the board is built using the basic gadgets modified to work for the Single-Step model. This includes changing any tilt path longer than the number of steps in a direction as shown in Figure \ref{fig:SSpath}. Similarly, the comparisons made in the \textsc{AND}, \textsc{OR}, and \textsc{Majority} need to be handled accordingly as well. An example of the Boolean sorting is shown in Figure \ref{fig:SSsort}. Each gate can check the output similar to the \textsc{Majority} check, which is shown in Figure \ref{fig:SSmaj}.  At each gate, it might take $O(w)$ cycles to walk the tiles to the sides of the gate to compute the output. Thus, the entire simulation takes $O(dw)$ total cycles.
\end{proof}


\section{Conclusion and Future Work}
We have shown that rotational tilt systems can efficiently simulate space-bounded, programmable Turing machines, which provides a framework for utilizing deterministic tilt systems as computational devices.  Through this, we have resolved the complexity of Relocation, Occupancy, Vacancy, and Reconfiguration as PSPACE-complete for deterministic tilt cycles for Full-Tilt with a 4-cycle and Single-Step with a cycle of length \SScycle. Further, we showed that the tape can be programmed directly or via tilts before computation begins. 

For efficiency, we show that a two-tape Turing machine can also be simulated in the Full-Tilt model with a constant number of cycles per operation. Single-Step has a harsh movement overhead based on the size of the tape. Thus, we show we can implement linear Systolic Arrays with to achieve some parallel efficiency improvement.
Finally, we show how Threshold Circuits (TC) can be efficiently implemented for any level of TC 
with both Full-Tilt and Single-Step. Full-Tilt only requires a number of cycles based on the depth, and Single-Step requires a number of cycles based on both the depth and width.
This work has introduced a number of interesting directions for future work.  

\vspace{-.2cm}

\begin{itemize}\setlength\itemsep{0em}
    
    \item For general tilt sequences, the reachability problems are PSPACE-complete even when restricted to singleton tiles \cite{Balanza:2020:SODA}. What is their complexity with deterministic tilt sequences and singleton tiles (no single domino)?

    \item What is the smallest Single-Step cycle where the four reachability-based problems are still PSPACE-complete? 
    What cycles are easy?

    \item What is the smallest Single-Step cycle that can directly simulate general computation? For both models, what are the fewest cycles needed to compute a given function? 
    
    
    \item We focused on rotational cycles, but~\cite{Becker2019} also considered 4-cycles with only 3 directions $\langle d, l, d, r\rangle$. Our \textsc{Majority} gate can be adapted to work in this model to give Threshold circuits with three directions. What other classes of circuits can be efficiently simulated? 
    

    \item It may be interesting to study how robust rotation-only systems are to perturbations in the rotation sequence.  A future work direction might be to design 
    rotation-based systems with built-in error correction that can withstand reasonable deviations from the intended rotation cycle.  Related questions might be to ask what the minimum alteration to a given tilt-sequence is to ensure some property, such as occupancy or vacancy.

\end{itemize}

\bibliographystyle{splncs04}
\bibliography{tiltCitations}


\end{document}